\documentclass[showkeys,preprintnumbers,amsmath,amssymb]{revtex4-2}

\usepackage[utf8]{inputenc}
\usepackage[pdftex]{graphicx}
\usepackage{caption}
\usepackage{subcaption}
\usepackage{mathrsfs}
\usepackage[colorlinks, breaklinks, urlcolor={blue}, linkcolor={red}, citecolor={blue}]{hyperref}
\usepackage{array}
\usepackage{amsmath}
\usepackage{type1cm}
\usepackage{lettrine}
\usepackage[english]{babel}
\usepackage{lmodern}
\usepackage{microtype}
\usepackage{booktabs}
\usepackage[T1]{fontenc}
\usepackage[boxed, vlined]{algorithm2e}
\usepackage{caption}
\usepackage{braket}
\usepackage{xcolor}
\usepackage{orcidlink}
\usepackage{bm}
\usepackage{bbold}
\usepackage{upgreek}
\DeclareMathOperator{\Tr}{Tr}

\begin{document}
	
	\title{Quantumness of electron transport in quantum dot devices through Leggett-Garg inequalities: A  non-equilibrium Green's function approach}
	
	\author{Thingujam Yaiphalemba Meitei}
	\affiliation{Department of Physics and Nanotechnology, Faculty of Engineering and Technology, SRM Institute of Science and Technology, Kattankulathur 603203, Tamil Nadu, India.}
	\author{Saikumar Krithivasan}
	\affiliation{Department of Physics and Nanotechnology, Faculty of Engineering and Technology, SRM Institute of Science and Technology, Kattankulathur 603203, Tamil Nadu, India.}
	\author{Arijit Sen \orcidlink{0000-0002-8624-9418}}%
	\email[]{arijits@srmist.edu.in}
	\affiliation{Department of Physics and Nanotechnology, Faculty of Engineering and Technology, SRM Institute of Science and Technology, Kattankulathur 603203, Tamil Nadu, India.}%

	\author{Md~Manirul Ali  \orcidlink{0000-0002-5076-7619}}
	\email[]{manirul@citchennai.net}
	\affiliation{ Centre for Quantum Science and Technology, Chennai Institute of Technology, Chennai 600069, India.
	}%
	
	\date{\today}







\begin{abstract}
Although coherent manipulation of electronic states can be achieved in quantum dot (QD) devices by harnessing nanofabrication tools, it is often hard to fathom the extent to which these nanoelectronic devices can behave quantum mechanically. Witnessing their nonclassical nature would thus remain of paramount importance in the emerging world of quantum technologies, since the coherent dynamics of electronic states plays there a crucial role. Against this backdrop, we resort to the general framework of Leggett-Garg inequalities (LGI) as it allows for distinguishing the classical and quantum transport through nanostructures by way of various two-time correlation functions. Using the local charge detection at two different time, we investigate here theoretically whether any quantum violation of the original LGI exists with varying device configurations and parameters under both Markovian and non-Markovian dynamics. Two-time correlators within LGI are derived in terms of the non-equilibrium Green's functions (NEGFs) by exactly solving the quantum Langevin equations. The present study of non-Markovian dynamics of quantum systems interacting with reservoirs is significant for understanding the relaxation phenomenon in the ultrafast transient regime to especially mimic what happens to high-speed quantum devices. We can potentially capture the effect of finite reservoir correlation time by accounting for level-broadening at the electrodes along with non-Markovian memory effects. Furthermore, the large bias restriction is no longer imposed in our calculations so that we can safely consider a finite bias between the electronic reservoirs. Our approach is likely to open up new possibilities of witnessing the quantumness for other quantum many-body systems as well that are driven out of the equilibrium.
\end{abstract}

\keywords{\textit{Quantum Dot Devices, Two-time Correlation Functions, Open Quantum Systems, Leggett-Garg Inequalities, Non-equilibrium Green's Functions, Quantum Transport}}
\pacs{05.60.Gg, 73.63.-b, 03.65.Yz}
\maketitle

\section{Introduction}\label{sec:Introduction}

Quantum nanostructure devices have attracted much attention in recent years  due to their potential applications in
the emerging quantum technologies where quantum coherence of electrons is the prime ingredient
\cite{heinrich2021quantum}. With the advent of quantum nanofabrication technology, it is now possible to design artificial
atoms and molecules using semiconductor quantum dots \cite{tarucha1996shell,oosterkamp1998microwave,fujisawa1998spontaneous}.
Coherent manipulation of electronic states can be achieved through single or double quantum dot devices
\cite{van2002electron,hayashi2003coherent,petta2005coherent,brandes2005coherent}. Probing nonclassical
or quantum nature of these nanodevices would remain of fundamental importance since quite often it is not clear to what extent the
system behaves quantum mechanically. The ability to distinguish between quantum and classical behavior plays
a crucial role in many emerging fields, such as quantum transport, quantum information processing,
quantum chemistry, quantum computing, quantum simulation, and sensing based on quantum materials or molecular nanosystems. Leggett-Garg inequalities (LGI) can provide a theoretical framework
\cite{leggett1985quantum} to distinguish between classical and quantum transport through quantum dot nanostructures,
which has been the main motivation of this work. The Leggett-Garg inequality is considered as the temporal analog \cite{paz1993proposed,ruskov2006signatures,palacios2010experimental,souza2011scattering,emary2012leggett} of
the Bell's inequality involving testable temporal correlation functions. Quantum systems manifest nonclassical correlations
through the violation of Leggett-Garg inequalities. The original motivation for these inequalities were to test the quantum
coherence in macroscopic systems \cite{leggett1985quantum,leggett2002testing}. The Leggett-Garg inequality can be
constructed as follows. Let us consider the measurement of an observable $Q(t)$ which is found to take up the values of
$+1$ or $-1$, whenever measured. One can then perform three set of experimental runs so that in the first set of runs, the
observable $Q(t)$ is measured at time $t_1$ and $t_2$; in the second run, $Q(t)$ is measured at $t_1$ and $t_3$; and in
the third run, $Q(t)$ is measured at $t_2$ and $t_3$. The two-time correlation functions
$C_{ji} = \langle Q(t_j) Q(t_i) \rangle$ can then be obtained by repeating such time-separated measurements.
Leggett-Garg imposed two classical assumptions: (a) measurement on classical systems reveal
well-defined pre-existing value, the measurement outcomes of the
observables $Q(t_1)$, $Q(t_2)$ and $Q(t_3)$ are predetermined prior to measurement and (b) any such
predetermined value can be measured without disturbing the system, implying that the measurement performed at one
time does not influence the subsequent dynamics of the system and the measurement outcomes at a later time. The above
classical assumptions imply the existence of a joint probability distribution \cite{leggett1985quantum,emary2014leggett,das2014unification,markiewicz2014unified}
to describe the time-separated measurement statistics of all three experimental runs. Classically, one can estimate the
average quantities $C_{ji}$ for these two-time measurements by this joint probability distribution. Subsequently, the Leggett-Garg
inequality \cite{leggett1985quantum} under those classical assumptions take the following form
\begin{eqnarray}
C_3 = C_{21} + C_{32} - C_{31} \le 1,
\label{LGI1}
\end{eqnarray}
where the detailed derivation of the LGI is discussed in Appendix-\ref{sec:appA}. Following the same arguments,
one can derive an LGI for measuring $Q(t)$ at four different time, $t_1$, $t_2$, $t_3$, and
$t_4$ resulting to the inequality
\begin{eqnarray}
C_4 = C_{21} + C_{32} + C_{43} - C_{41} \le 2.
\label{LGI2}
\end{eqnarray}
Quantum mechanically, the average values of this type of two-time measurements can be obtained
\cite{fritz2010quantum,emary2014leggett,chen2014investigating,ali2017probing} through the
expectation values of the symmetrized Hermitian operator $(Q(t_j) Q(t_i) + Q(t_i) Q(t_j))/2$.
Violation of Leggett-Garg inequality implies either the absence of a classical realistic description of the system or the
impossibility of measuring the system without disturbing it, quantum systems can violate the inequalities on both ground.
Experimental violation of LGI is demonstrated in diverse range of physical systems, for example, superconducting
qubit \cite{palacios2010experimental,knee2016strict,santini2022experimental}, photonic systems \cite{xu2011experimental,dressel2011experimental,goggin2011violation,suzuki2012violation},
spin systems \cite{athalye2011investigation,souza2011scattering,moreira2015modeling,medina2021quantum},
phosphorus impurities in silicon \cite{knee2012violation},
and nitrogen-vacancy defect in diamond \cite{waldherr2011violation}. Quantum violations of LGI have been studied
theoretically in optomechanical system \cite{lambert2011macrorealism}, atomic ensemble \cite{budroni2015quantum},
oscillating neutral kaons and neutrino oscillations \cite{gangopadhyay2013probing,formaggio2016violation}, and
even in biological light-harvesting protein complex \cite{wilde2010could,li2012witnessing}. LGI violation is used as
an indicator/witness of nonclassicality for open quantum systems \cite{emary2014leggett,chen2014investigating,ali2017probing}.
In this work, we use LGI violation as a tool to probe ``quantumness'' for electron transport through double quantum dots,
the experimental violation of LGI would then exclude the possibility of a classical description of transport through the
nanostructure. We consider a nanosystem of two laterally coupled single-level quantum dots coupled to two electrodes, and
also, a parallel configuration when each dot is coupled to both the left and right electrodes.
Recent experimental investigations on double quantum dot systems
\cite{kiesslich2007noise,shinkai2009correlated,kagan2015charge,dorsch2021heat} provide an extra motivation to study
the LGI violation in such systems. Quantum dot systems in presence of the electronic reservoirs are considered
as open quantum systems, and non-equilibrium transport through these nanostructures are often studied using quantum master
equation approach \cite{MatisseTu2008non,jin2010non,tu2012transient}. The open system dynamics has been the subject of interest
for many studies in presence of environmental noise \cite{carmichael1999statistical,breuer2002theory,gardiner2004quantum}.
The resulting dissipation and decoherence dynamics lead to the loss of quantumness of the system. From the perspective of
probing quantumness, the LGI violation of the open system is mainly studied under Born-Markov approximation justifying
weak coupling, wide band limit, and short correlation time of the reservoir. It is relatively easy
to evaluate the two-time correlation functions when Born-Markov approximation is valid, two-time correlation functions
can then be calculated using the quantum regression theorem where the memory effect is totally ignored. However, non-Markovian
dynamics of quantum system interacting with environment is significant to model the relaxation phenomenon in the short
time transient regime, applicable for high speed quantum devices. In the present work, we use Heisenberg equation of
motion approach to obtain the exact dynamics of the two-time correlation functions in terms of nonequilibrium Green's
functions. Thus, our results are applicable to both Markov and non-Markovian regime.

Recently, an extended LG-inequality (ELGI) has been developed and investigated for electron transport through
nanostructures, under the classical assumption that measurements can be performed non-invasively, and also, under classical
Markov process, based on the Chapman-Kolmogorov equation in stochastic theory \cite{lambert2010distinguishing}.
The ELGI is claimed to mimic the original LG-inequality when the initial zero-time state is considered as the steady state described by the stationary density matrix for the system, and the measurements are performed non-invasively. Moreover, they assumed that the DQD system is weakly coupled to the electrodes, also assumed a large bias condition such that higher-order tunneling, level-broadening, and non-Markovian effects can be completely neglected \cite{gurvitz1996microscopic,gurvitz1998rate}. Assuming weak coupling, large bias, and Coulomb blockade, two-time correlation functions were calculated with respect to a stationary density matrix of the system, where the time evolution of the observable is obtained through a Liouvillian superoperator under Born-Markov Lindblad master equation. For localized charge detection, violation of the {\it extended} LGI is shown in the short-time transient regime \cite{lambert2010distinguishing}. It is important to note that the Markov dynamics is unable to capture the short-time transient dynamics of the observable, for which reservoir's memory effect is completely ignored. Contrary to that, our approach can probe the Leggett-Grag inequalities in the full system-reservoir parameter regime.

\section{Electronic Transport and System Dynamics}

\begin{figure}[htp]
\centering
\begin{subfigure}[b]{0.45\textwidth}
\centering
\includegraphics[width=\textwidth]{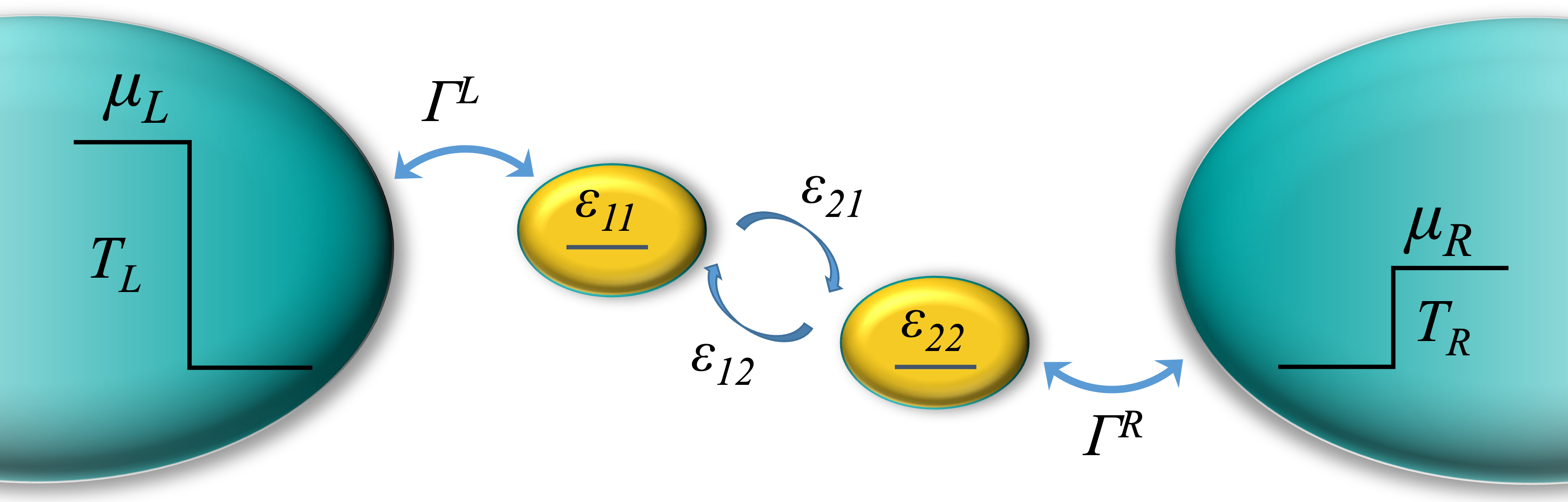}
\caption{Quantum dots in series}
\label{fig1a}
\end{subfigure}
\vskip 0.2cm
\begin{subfigure}[b]{0.45\textwidth}
\centering
\includegraphics[width=\textwidth]{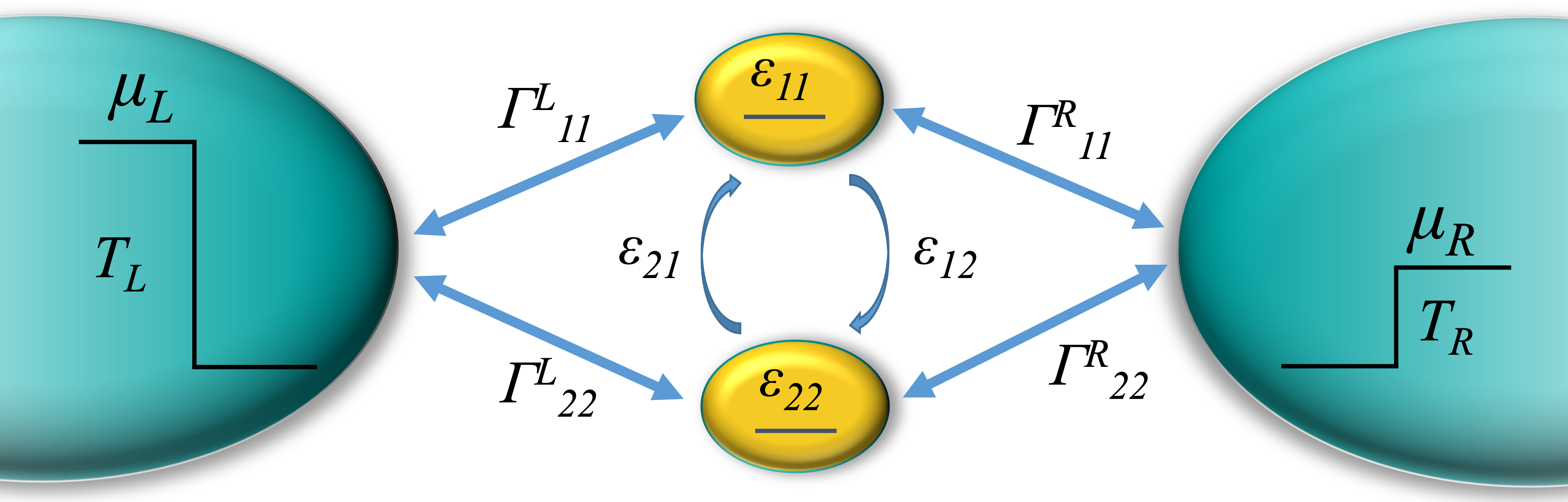}
\caption{Quantum dots in parallel}
\label{fig1b}
\end{subfigure}
\caption{Schematic diagram of double quantum dot nanostructures with two configurations (a) first dot is coupled to
the left electrode while the second dot is coupled to the right electrode, where $\Gamma^{L,R}$ describe the left/right
tunneling rates (b) both dots are coupled to the electrodes with tunneling rates $\Gamma^{L,R}_{11}$ and
$\Gamma^{L,R}_{22}$. The dot energy levels are $\epsilon_{11}$ and $\epsilon_{22}$, and $\epsilon_{12}$ represents
the inter dot tunneling amplitude. The electrodes are taken initially in thermal equilibrium with temperatures $T_{L,R}$
and chemical potentials $\mu_{L,R}$. Local charge measurements are done for the second dot occupation.}
\label{fig1ab}
\end{figure}
In this work, we consider a system of double quantum dot interacting with two fermionic reservoirs maintained at finite
chemical potential and at finite temperatures. The total Hamiltonian of the system in presence of the electronic reservoirs is given by (see Fig.~\ref{fig1ab})
\begin{eqnarray}
H = H_{DQD} + H_E + H_I,
\label{htot}
\end{eqnarray}
where $H_{DQD}$ is the Hamiltonian of two single-level quantum dots
\begin{eqnarray}
H_{DQD} = \sum_{i,j=1}^{2} \epsilon_{ij}  a_i^{\dag}  a_j,
\label{hdqd}
\end{eqnarray}
with $a_i$ and $a_i^{\dag}$ being the fermionic annihilation and creation operators associated to the $i$th
quantum dot, $\epsilon_{ii}$ represents the energy level of the $i^{th}$ QD, and $\epsilon_{ij}$ with $i \ne j$ is the
tunnel coupling between the two dots. The Hamiltonian of the two electronic reservoirs (electrodes)
\begin{eqnarray}
H_E = \sum_{\alpha=L,R} \sum_{k} \epsilon_{\alpha k}  c_{\alpha k }^{\dag}  c_{\alpha k },
\label{hlead}
\end{eqnarray}
where the label $\alpha$ denotes the left or the right fermionic electrode, the left dot is coupled to the left electrode while
the right dot is coupled to the right electrode, and $c_{\alpha k }^{\dag}$ ($c_{\alpha k }$) is the creation (annihilation)
operator of the $k^{th}$ level in electrode $\alpha$. The Hamiltonian describing the coupling between the double dot system
and the electrodes
\begin{eqnarray}
H_I = \sum_{i \alpha k} (V_{i\alpha k}  a_i^{\dag}  c_{\alpha k } + V_{i\alpha k}^*  c_{\alpha k }^{\dag}  a_i  ),
\label{hint}
\end{eqnarray}
where $V_{i\alpha k}$ is the coupling strength of $k^{th}$ level of reservoir $\alpha$ to a particular dot level $i$. The same Hamiltonian (\ref{htot}) can also describe the situation where quantum dots are in parllel configuration (see Fig.~\ref{fig1b}).
We investigate the dynamics of the Leggett-Garg inequalities (\ref{LGI1}) and (\ref{LGI2}) for the DQD system
with the measurement operator $Q(t)=2 n_2(t)  - 1$, where $n_2(t)=a_2^\dag(t)   a_2(t)$ is the occupation number
operator in the second quantum dot at time $t$. The particle number of the second dot can be measured by a localized
charge detector. In the context of Leggett-Garg inequality, local charge measurements at different time is considered
for closed system under continuous weak measurements \cite{korotkov2001output,ruskov2006signatures,williams2008weak}.
In contrast, we consider here strong projective noncontinuous measurements in open transport scenario.
The measurement outcomes of the observable $Q(t)$ take dichotomic values $\pm 1$ corresponding to the situations
when the second dot is occupied ($n_2(t)=1$) or empty ($n_2(t)=0$). Leggett-Garg inequalities provide classical
bounds to the quantities $C_3$ and $C_4$, and we probe the inequalities (\ref{LGI1}) and (\ref{LGI2}) for electron
transport through quantum dot nanostructure with time-separated measurements of the observable $Q(t)$.

\section{Quantumness through LGI for an isolated quantum dot system}

Before we get into the open system scenario, it is worth exploring LGI for the closed double quantum dot system. The system of our interest is just a quantum double dot at two different energy levels governed by the Hamiltonian $H_{DQD} = \sum_{i,j=1}^{2}\epsilon_{ij}  a_i^{\dag}  a_j$, where $ a_i^{\dag}$ and $ a_j$ are fermionic creation and annihilation operators, and \textit{i,j} are labels which would take values of either 1 or 2, referring to the first and second dot with respective energies of $\epsilon_{11}$ and $\epsilon_{22}$. The inter-dot coupling strengths are represented by $\epsilon_{12}$ and $\epsilon_{21}$. It may be noted that $\epsilon_{11}$ and $\epsilon_{22}$ can be an arbitrary real number, while $\epsilon_{12}$ and $\epsilon_{21}$ are complex conjugate of each other so that the Hamiltonian would in turn be hermitian. The creation and annihilation operators $a_i^\dag(t)$ and $a_i(t)$ obey the fermionic anti-commutation relations. We use the Heisenberg's equation of motion approach in evaluating the time evolution of $a_i(t)$

\begin{eqnarray}
	\label{aidiff}
	\frac{d}{dt}  a_i (t) = -i \left[ a_i (t), H_{DQD} \right] = -i  \sum_{j} \epsilon_{ij}  a_j (t).
\end{eqnarray}

\begin{figure*}[htp]
	\centering
	\begin{subfigure}[b]{0.45\textwidth}
		\centering
		\includegraphics[width=\textwidth]{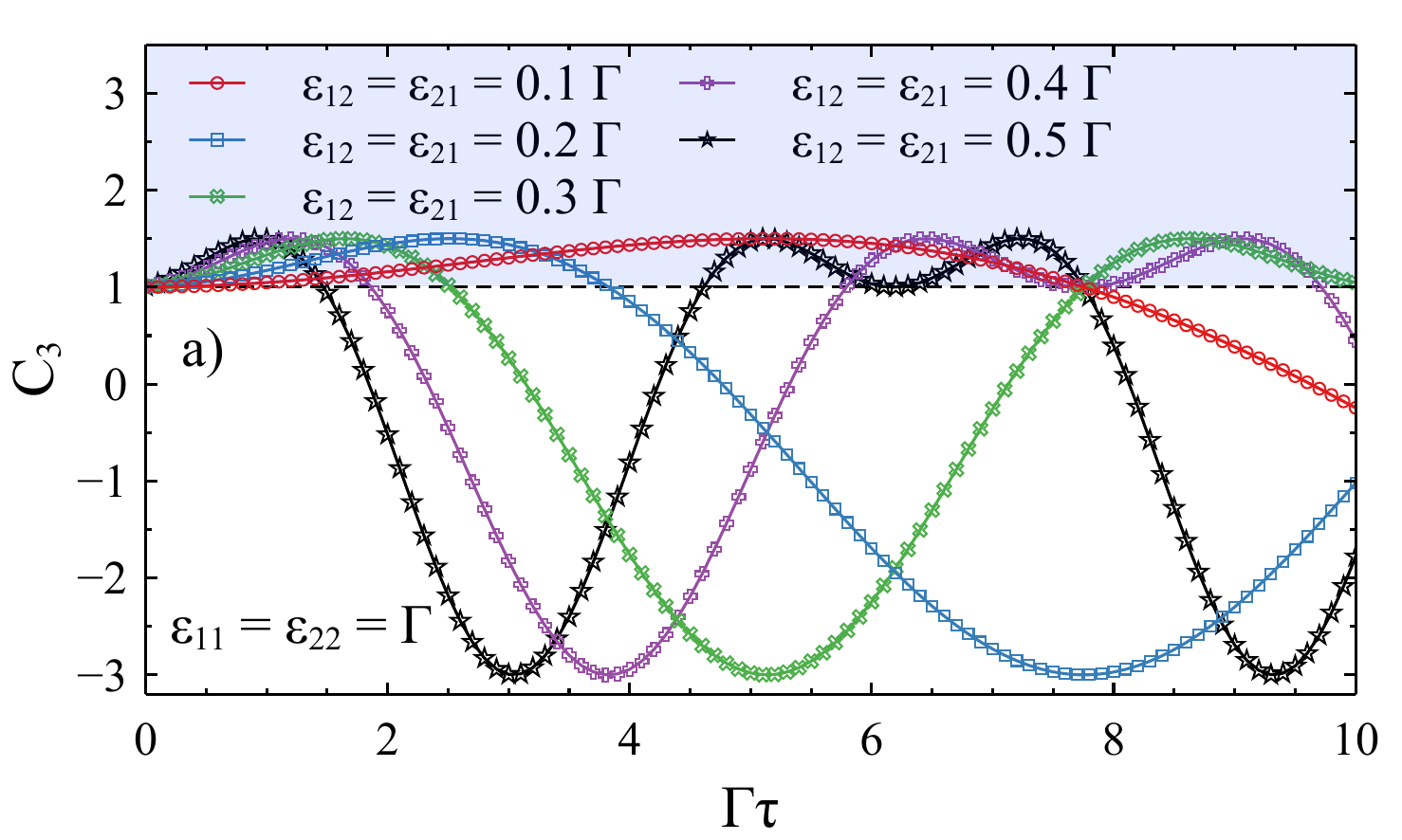}
	\end{subfigure}
	\hskip -0.0cm
	\begin{subfigure}[b]{0.45\textwidth}
		\centering
		\includegraphics[width=\textwidth]{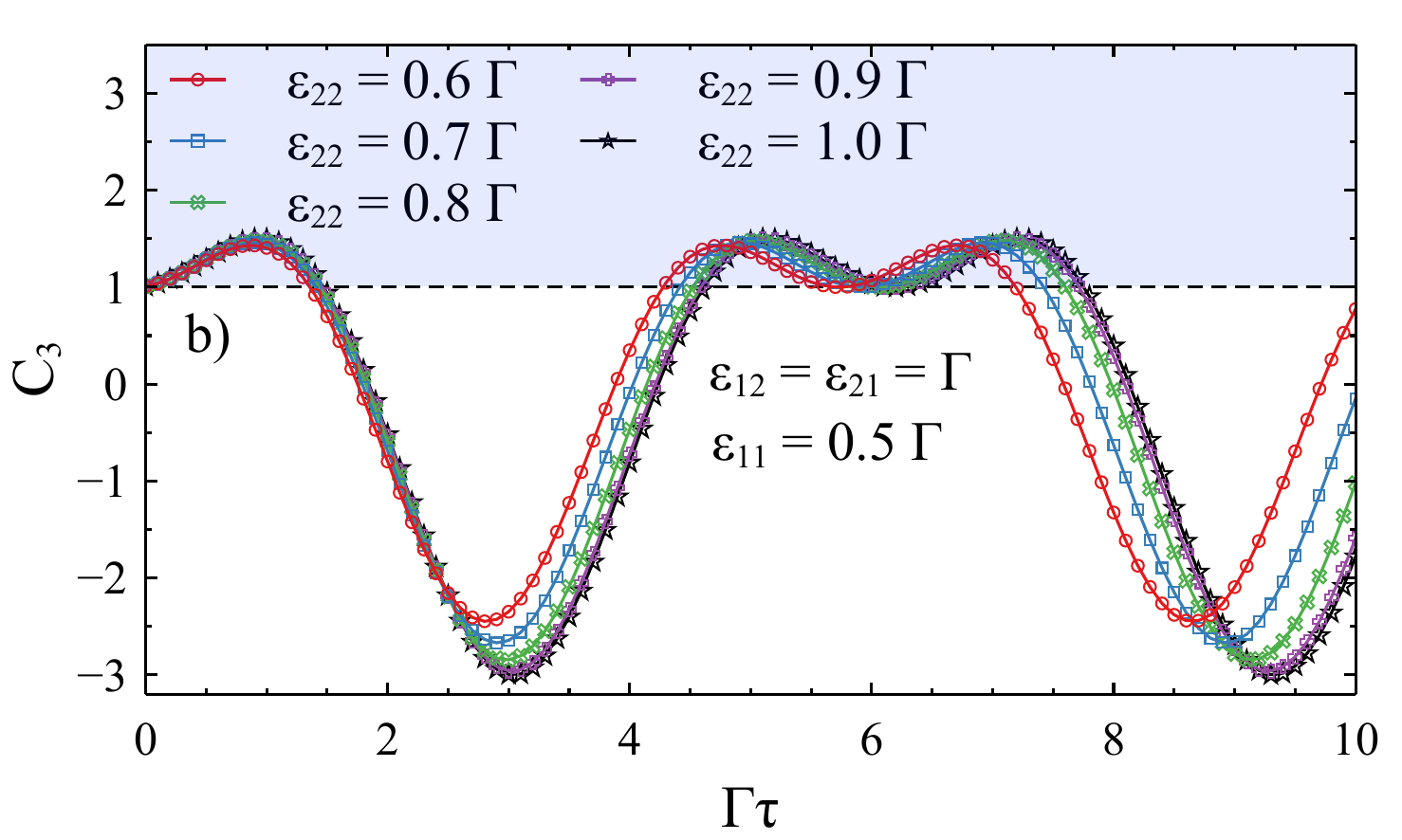}
	\end{subfigure}
	\caption{We show the dynamics of correlator $C_3$ for double quantum dot
		closed system in absence of the electronic reservoirs by (a) varying inter dot coupling $\epsilon_{12}$
		with fixed values of dot energies $\epsilon_{11}=\epsilon_{22}=\Gamma$ (b) varying the on-site energy $\epsilon_{22}$ of the
		second dot with fixed values of $\epsilon_{11}=\Gamma$ and $\epsilon_{12}=0.5\Gamma$.}
	\label{fig2}
\end{figure*}

\begin{figure*}[htp]
	\centering
	\begin{subfigure}[b]{0.45\textwidth}
		\centering
		\includegraphics[width=\textwidth]{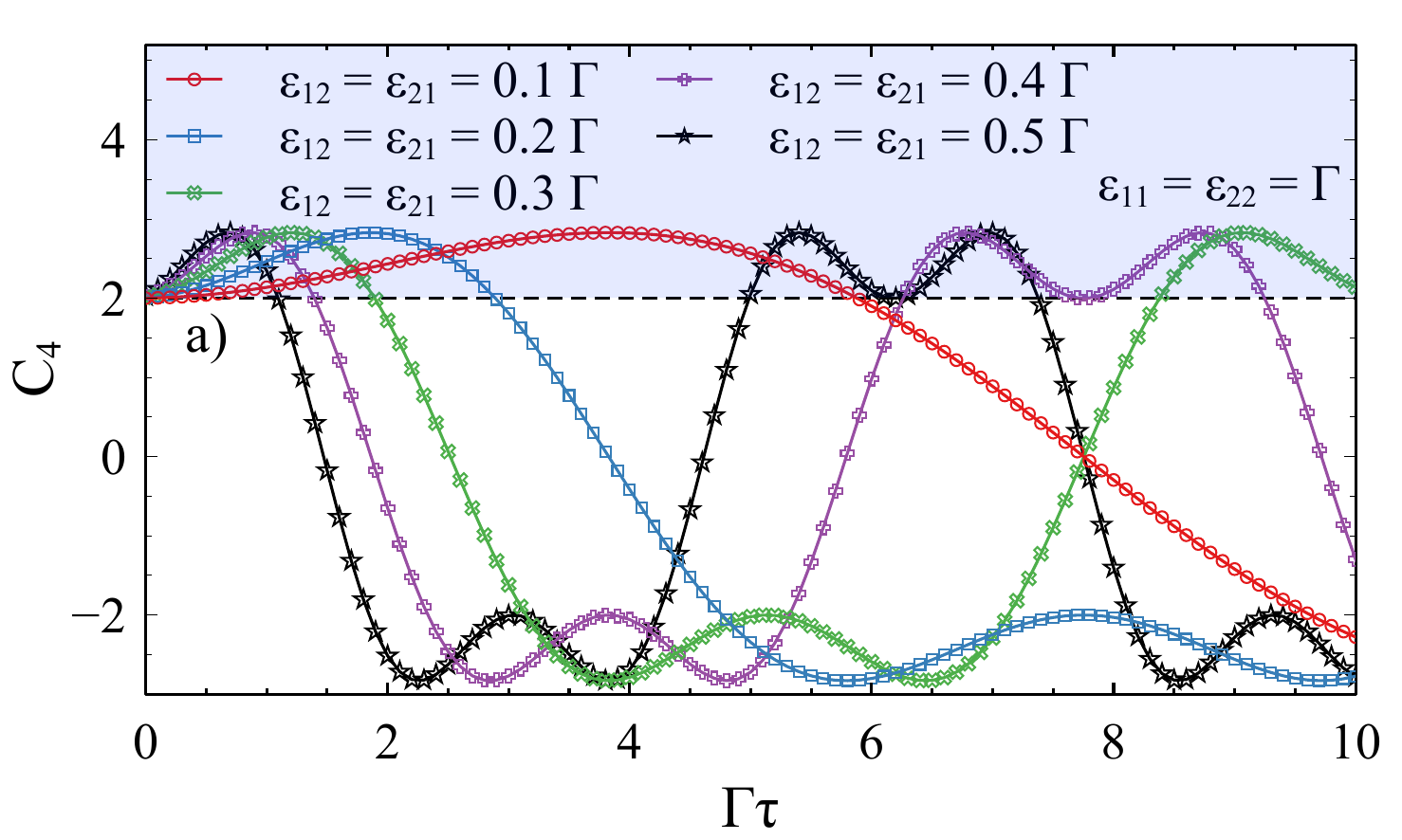}
	\end{subfigure}
	\hskip -0.0cm
	\begin{subfigure}[b]{0.45\textwidth}
		\centering
		\includegraphics[width=\textwidth]{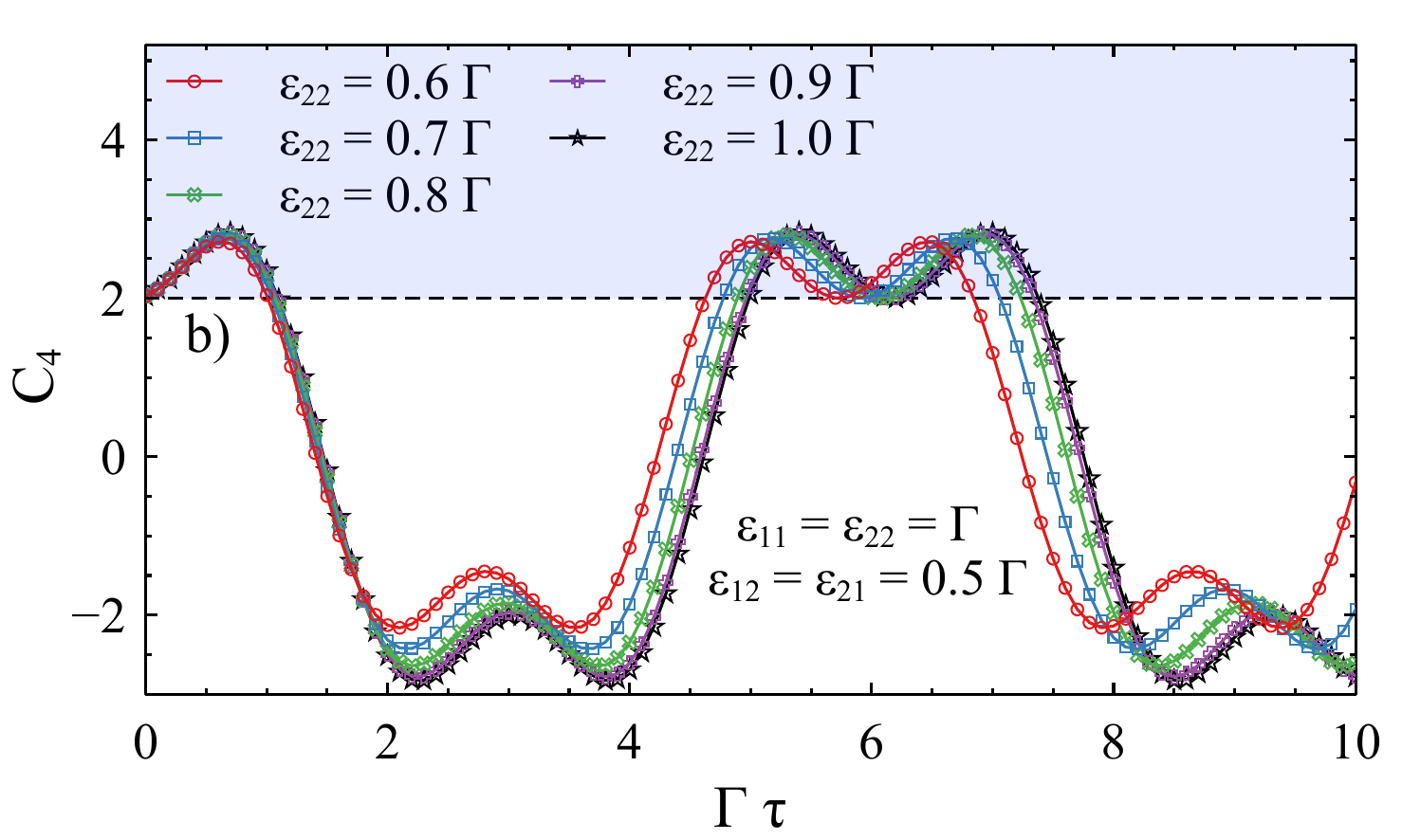}
	\end{subfigure}
	\caption{We show the dynamics of LGI correlator $C_4$ for double quantum dot
		closed system in absence of the electronic reservoirs by (a) varying inter dot coupling $\epsilon_{12}$
		with fixed values of dot energies $\epsilon_{11}=\epsilon_{22}=\Gamma$ (b) varying the on-site energy $\epsilon_{22}$ of the
		second dot with fixed values of $\epsilon_{11}=\Gamma$ and $\epsilon_{12}=0.5\Gamma$.}
	\label{fig3}
\end{figure*}

Considering the linearity of the equation (\ref{aidiff}), one can express the solution to the above equation as
\begin{eqnarray}
	\label{aisol}
	a_i (t) = \sum_{j} w_{ij}(t,t_0) a_j (t_0),
\end{eqnarray}
from which we have the following equation for $w_{ij}(t,t_0)$
\begin{eqnarray}
	\label{wij}
	\frac{d}{dt}  w_{ij}(t,t_0) =  \sum_{m} \epsilon_{im}  w_{mj}(t,t_0).
\end{eqnarray}
For this closed DQD, the two-time correlation function
$\langle n_2(t_2)  n_2(t_1) \rangle$ can be expressed in terms of the functions $w_{ij}(t,t_0)$ as
\begin{eqnarray}
	\label{n2n1C}
	\nonumber
	\langle n_2(t_2)  n_2(t_1) \rangle &=&\langle a_2^{\dag}(t_2)  a_2(t_2) a_2^{\dag}(t_1)  a_2(t_1) \rangle \\
	&= & \dfrac{1}{2} \big( \big|w_{21}^* (t_2)\big|^2 \big|w_{21}^* (t_1)\big|^2 + \big|w_{21}^* (t_2)\big|^2  w_{21}^{\ast}(t_1) w_{22}(t_1) + \nonumber \\
	&& w_{21}^{\ast}(t_2)w_{22}(t_2)w_{22}^{\ast}(t_1)w_{21}(t_1)+w_{21}^{\ast}(t_2)w_{22}(t_2)\big|w_{22}^*(t_1)\big|^2 + \nonumber\\
	&& w_{22}^{\ast}(t_2)w_{21}(t_2)\big|w_{21}^*(t_1)\big|^2 + w_{22}^{\ast}(t_2)w_{21}(t_2)w_{21}^{\ast}(t_1)w_{22}(t_1) +	\nonumber\\
	&& \big|w_{22}(t_2)\big|^2 w_{22}^{\ast}(t_1) w_{21}(t_1)   + \big|w_{22}(t_2)\big|^2 \big|w_{22}(t_1)\big|^2 \big).
\end{eqnarray}

The initial state of the DQD system is taken arbitrarily as $\dfrac{1}{\sqrt{2}} \big(\big|01\rangle + \big|10\rangle)$, where $\big|01\rangle $ represents a state with first dot unoccupied and the second dot occupied. The two-time correlation function of the observable $Q(t)$ given by
\begin{eqnarray}
	\nonumber
	\langle Q(t_2)  Q(t_1) \rangle &=&4 \langle n_2(t_2)  n_2(t_1) \rangle - 2 \langle n_2(t_2) \rangle - 2 \langle n_2(t_1) \rangle + 1 \\
	\nonumber
	&= & 2 \big( \big|w_{21}^* (t_2)\big|^2 \big|w_{21}^* (t_1)\big|^2 + \big|w_{21}^* (t_2)\big|^2  w_{21}^{\ast}(t_1) w_{22}(t_1) +  \\
	\nonumber
	&& w_{21}^{\ast}(t_2)w_{22}(t_2)w_{22}^{\ast}(t_1)w_{21}(t_1)+w_{21}^{\ast}(t_2)w_{22}(t_2)\big|w_{22}^*(t_1)\big|^2 + \\
	\nonumber
	&& w_{22}^{\ast}(t_2)w_{21}(t_2)\big|w_{21}^*(t_1)\big|^2 + w_{22}^{\ast}(t_2)w_{21}(t_2)w_{21}^{\ast}(t_1)w_{22}(t_1) +\\
	\nonumber
	&& \big|w_{22}(t_2)\big|^2 w_{22}^{\ast}(t_1) w_{21}(t_1)   + \big|w_{22}(t_2)\big|^2 \big|w_{22}(t_1)\big|^2 \big) - \\
	\nonumber
	&& \big( \big|w_{21}(t_2)\big|^2 + w_{21}^{\ast}(t_2) w_{22}(t_2) + w_{22}^{\ast}(t_2) w_{21}(t_2) +  \big|w_{22}(t_2)\big|^2 \big) -\\
	&& \big( \big|w_{21}(t_1)\big|^2 + w_{21}^{\ast}(t_1) w_{22}(t_1) + w_{22}^{\ast}(t_1) w_{21}(t_1) +  \big|w_{22}(t_1)\big|^2 \big) +1.
	\label{C21c}
\end{eqnarray}

By solving equation (\ref{wij}) with the initial condition $w_{ij}(t_0,t_0)=\delta_{ij}$,
one can investigate the dynamical behaviour of the Leggett-Garg inequalities (\ref{LGI1}) and (\ref{LGI2})
for this closed DQD system. The analytic solutions for $w_{ij}(t,t_0)$ are given in the Appendix-\ref{sec:appB}.
We take the measurement operator $Q(t)=2 n_2(t)  - 1$, where the time-dependent particle
number operator associated to the second dot is $n_2(t)=a_2^\dag(t) a_2(t)$. Then considering charge detection
at four different times $t_1$,  $t_2$, $t_3$, and $t_4$, one can calculate the LGI temporal correlation terms $C_{21}$,
$C_{32}$, $C_{31}$, $C_{43}$, and $C_{41}$ for this double quantum dot system to finally
estimate the values of $C_3$ and $C_4$. In Fig.~\ref{fig2}, we show the dynamics of Leggett-Garg
inequality correlator $C_3$ for DQD closed system in absence of the electronic
reservoirs. The dynamics of $C_3$ is shown in Fig.~\ref{fig2}a with varying inter dot coupling $\epsilon_{12}$ when
the dot energy levels are fixed as $\epsilon_{11}$$=\epsilon_{22}=\Gamma$. For this closed DQD system,
the violations of LGI occurs periodically with the time interval between two measurements take a value
$\tau=2n\pi/\sqrt{(\epsilon_{11}-\epsilon_{22})^{2}+4\epsilon^{2}_{12}}$.
From the expression of $\tau$, it is seen that time period of violation depends inversely on the difference
between the dot energy levels and also on the interdot coupling strength. Hence the frequency of the
violation decreases as the inter dot coupling is reduced.This periodicity matches with the revival time i.e,
the time after which the system returns to the same state.
In Fig.~\ref{fig2}b, we demonstrate $C_3$ dynamics by varying the on-site energy $\epsilon_{22}$
of the second dot.  We take a value of $\epsilon_{11}=\Gamma$ and the inter dot coupling is fixed at
$\epsilon_{12}=0.5\Gamma$. This shows that the periodic violations of LGI are less sensitive to the dot
energies compared to inter dot coupling. Similar dynamical violation of Leggett-Grag inequality is shown through the correlator $C_4$ in Figs.~\ref{fig3}a and  \ref{fig3}b. We also observed that the dynamical characteristics of the Leggett-Garg inequalities do not
depend significantly on the choice of the initial states.

\section{Quantumness of quantum dot device attached to fermionic reservoirs}

We consider a central system (double quantum dot) coupled to fermionic reservoirs as dictacted by the total Hamiltonian
given in equation (\ref{htot}). Since the central system is connected to the reserviors, it can exchange both particle and
energy with the reserviors, due to which the particle number and energy of the central system is not conserved, and thus
making the dynamics of central system non unitary. Instead of using Born-Markov Lindblad master equation to obtain the
dynamics of the central system, which works only in Markov and weak coupling regime, we use quantum Langevin equation
approach which enable us to obtain the time dynamics exactly without the weak couling and Markov approximations.
The time evolution of the operators $a_i(t)$ and $ c_{\alpha k}(t)$ are given by the Heisenberg equations of motion
\begin{eqnarray}
\label{Heq1}
\frac{d}{dt} a_i (t) =  -i  \sum_{j} \epsilon_{ij}  a_j (t) -i \sum_{\alpha k } V_{i\alpha k}  c_{\alpha k } (t), \\
\frac{d}{dt} c_{\alpha k} (t) = -i  \epsilon_{\alpha k} c_{\alpha k} (t) -i \sum_{i} V_{i\alpha k}^\ast  a_{i} (t).
\label{Heq2}
\end{eqnarray}
Solving the equation of motion (\ref{Heq2}) we get the time-dependent operator $c_{\alpha k} (t)$ of
the electrode $\alpha$ as
\begin{eqnarray}
c_{\alpha k} (t)  = c_{\alpha k} (t_0) e^{-i \epsilon_{\alpha k} (t-t_0)}
- i \sum_{i} \int_{t_0}^t d\tau  V_{i\alpha k}^\ast a_i (\tau) e^{-i \epsilon_{\alpha k} (t-\tau)}.
\label{cbath}
\end{eqnarray}
Substituting the solution (\ref{cbath}) in (\ref{Heq1}) we arrive at the following quantum
Langevin equation
\begin{eqnarray}
\frac{d}{dt}  a_i (t) = -i  \sum_{j}\epsilon_{ij}  a_j (t) -  \sum_{\alpha j} \int_{t_0}^t d \tau  g_{ \alpha ij }(t,\tau)  a_j (\tau) - i \sum_{\alpha k } V_{i\alpha k}  c_{\alpha k } (t_0) e^{-i \epsilon_{\alpha k} (t-t_0)}.
\label{Langevin}
\end{eqnarray}
The first term in the right of quantum Langevin equation (\ref{Langevin}) is determined by the central system (quantum dots)  of the
nanostructure, second term describes the dissipation caused by the coupling to the electrodes, and the last term represents
the fluctuation induced by the fermionic environment (electrodes). Here $g_{\alpha ij}(t,\tau)$ represents the memory
kernel and is given by
\begin{eqnarray}
\label{g01}
g_{ \alpha ij }(t,\tau) = \sum_{k \in \alpha} V_{i \alpha k } V_{j \alpha k }^\ast  e^{-i \epsilon_{\alpha k} (t-\tau)}.
\end{eqnarray}
In the continuum limit, the memory kernel can be written as $g_{ \alpha ij }(t,\tau)$
$=\int \frac{d\omega}{2\pi} ~ J_{\alpha i j}(\epsilon)  e^{-i \epsilon (t-\tau)}$, where
$J_{\alpha i j}(\epsilon)$ $= 2\pi \sum_{k \in \alpha} V_{i \alpha k} V_{j \alpha k }^\ast \delta(\epsilon-\epsilon_{\alpha k})$
is the spectral density (level broadening) which encode the interaction between the dots and the electrodes. The integral kernel
$g_{ \alpha ij }(t,\tau)$ characterizes all the non-Markovian memory effects of the electronic reservoirs on the central dots.
Because of the linearity of equation (\ref{Langevin}), the general solution to the quantum Langevin equation can be expressed as
\begin{eqnarray}
\label{aidot}
 a_i (t) = \sum_j u_{ij} (t, t_0)  a_j (t_0) +  F_i (t),
\end{eqnarray}
where $u_{ij} (t,t_0)=\langle \{ a_i (t) , a_j^{\dag}(t_0) \} \rangle$ is the retarded Green function in
Keldysh formalism of nonequilibrium quantum transport theory \cite{keldysh1965diagram,jin2010non}.
The second term $F_i (t)$ is the noise operator, and we don't assume correlation of the noise operator at different
instants of time to be delta correlated. This enable us to capture the non-Markovian memory effect in the dynamics
of the system. By substituting the solution (\ref{aidot}) in equation (\ref{Langevin}) one can obtain the following
differential equations governing the time dynamics of $u_{ij} (t,t_0)$ and $F_i (t)$ as follows
\begin{align}
\nonumber
\frac{d}{dt} u_{ij} (t, t_0) & + i \sum_m \epsilon_{im} u_{mj} (t,t_0) \\
& + \sum_{\alpha} \int_{t_0}^t d\tau \sum_m g_{\alpha i m}(t,\tau) u_{m j} (\tau, t_0) = 0,
\label{uij}
\end{align}
\begin{eqnarray}
\nonumber
\frac{d}{dt} F_{i} (t) &+& i \sum_m \epsilon_{i m} F_{m}(t) + \sum_{\alpha m} \int_{t_0}^t d\tau
g_{\alpha i m} (t,\tau) F_m (\tau)  \\
\label{noise}
&&{} = - i \sum_{\alpha k} V_{i\alpha k}  c_{\alpha k} (t_0) e^{-i \epsilon_{\alpha k} (t-t_0) }.
\end{eqnarray}

\subsection{Noise operator of fermionic reservoirs for non-equilibrium electronic transport}

The analytic solution of the noise operator $F_{i}(t)$ can be obtained by soving the inhomogeneous
equation (\ref{noise}) with the initial condition $F_{i}(t_0)=0$ as we assume initially there is no interaction between the central system and the fermionic reservoirs. The solution for equation(\ref{noise}) is given by
\begin{eqnarray}
	F_i(t)\!=\!- i \sum_{j \alpha k} \int_{t_0}^{t} d\tau u_{ij}(t,\tau) V_{j \alpha k} c_{\alpha k} (t_0) e^{-i \epsilon_{\alpha k} (\tau-t_0)}.
	\label{noise2}
\end{eqnarray}
\vskip -0.2cm
We assume that the double quantum dot system is uncorrelated with the reservoirs at the initial time $t=0$. The initial state of the
total system is considered to be a product state, in which the system is in an arbitrary state $\rho_s(t_0)$, and the reservoirs are
initially in thermal equilibrium as follows
\vskip -0.3cm
\begin{eqnarray}
	\rho_{tot}(t_0) = \rho_s(t_0) \prod_{\alpha} \rho_{\alpha} (t_0),
\end{eqnarray}
\vskip -0.4cm
where
\vskip -0.4cm
\begin{eqnarray}
	\rho_{\alpha} (t_0) = \frac{\exp \big[- \beta_\alpha (H_\alpha - \mu_\alpha N_\alpha ) \big]}
	{\Tr \exp \big[- \beta_\alpha (H_\alpha - \mu_\alpha N_\alpha ) \big]}.
\end{eqnarray}
Here $\mu_\alpha$ is the chemical potential of $\alpha^{th}$ electrode, $\beta_\alpha=1/(k_B T_\alpha)$ is the inverse temperature
of electrode $\alpha$ at initial time $t_0$, and $N_\alpha=\sum_k c_{\alpha k }^{\dag} c_{\alpha k}$ is the
total particle number for the electrode $\alpha$. Then using the solution (\ref{noise2}) one can obtain the two-time
noise correlation functions given by
\begin{eqnarray}
	\label{vij}
	\nonumber
	&& \!\!\!\!\! \langle F_{j}^\dagger(t_2) F_{i} (t_1) \rangle = v_{ij}(t_1,t_2) \\
	\nonumber
	&&{} \!\!\!\!\!= \sum_{\alpha m n} \int_{t_0}^{t_1} \!\!\!\!d\tau_1  \int_{t_0}^{t_2} \!\!\!\!d\tau_2~ u_{i m} (t_1,\tau_1)
	{\widetilde{g}}_{\alpha m n} (\tau_1,\tau_2) u_{j n}^{\ast}(t_2,\tau_2) \\
	&&{}\!\!\!\!\!= \sum_{\alpha} \int_{t_0}^{t_1} \!\!\!\!d\tau_1 \int_{t_0}^{t_2} \!\!\!\!d\tau_2 \Big[ {\bf u}(t_1,\tau_1)
	{\widetilde{\bf g}}_{\alpha} (\tau_1,\tau_2) {\bf u}^{\dag}(t_2,\tau_2) \Big]_{ij}~,
\end{eqnarray}
\vskip -0.2cm
and
\vskip -0.2cm
\begin{eqnarray}
	\label{vijbar}
	\nonumber
	&& \!\!\!\!\! \langle F_{i} (t_1) F_{j}^\dagger(t_2)  \rangle = {\overline v}_{ij}(t_1,t_2) \\
	\nonumber
	&&{} \!\!\!\!\!= \sum_{\alpha m n} \int_{t_0}^{t_1} \!\!\!\!d\tau_1  \int_{t_0}^{t_2} \!\!\!\!d\tau_2~ u_{i m} (t_1,\tau_1)
	{\overline{g}}_{\alpha m n} (\tau_1,\tau_2) u_{j n}^{\ast}(t_2,\tau_2) \\
	&&{}\!\!\!\!\!= \sum_{\alpha} \int_{t_0}^{t_1} \!\!\!\!d\tau_1 \int_{t_0}^{t_2} \!\!\!\!d\tau_2 \Big[ {\bf u}(t_1,\tau_1)
	{\overline{\bf g}}_{\alpha} (\tau_1,\tau_2) {\bf u}^{\dag}(t_2,\tau_2) \Big]_{ij}~.
\end{eqnarray}
The time correlation functions are as under:
\begin{eqnarray}
	\label{gtilde}
	{\widetilde{g}}_{\alpha m n} (\tau_1,\tau_2) = \sum_k V_{m \alpha k } V_{n \alpha k }^\ast f_{\alpha}(\epsilon_{\alpha k})
	e^{-i \epsilon_{\alpha k} (\tau_1-\tau_2)},
\end{eqnarray}
\vskip -0.5cm
\begin{eqnarray}
	\label{gtbar}
	{\overline{g}}_{\alpha m n}\!(\tau_1,\tau_2) \!=\!\!\! \sum_k \!V_{m \alpha k } V_{n \alpha k }^\ast (1\!\!-\!\!f_{\alpha}(\epsilon_{\alpha k}))
	e^{-i \epsilon_{\alpha k} (\tau_1-\tau_2)},
\end{eqnarray}
where $f_{\alpha}(\epsilon_{\alpha k})$ $=\langle c_{\alpha k}^{\dagger} (t_0)  c_{\alpha k} (t_0) \rangle$.
The function $v_{ij}(t_1,t_2)$ is related to the lesser Green function in Keldysh formalism \cite{jin2010non}.
In the continuum limit, the time correlation functions $g_{ \alpha ij }(t,\tau)$,
${\widetilde{g}}_{\alpha m n} (\tau_1,\tau_2)$, and ${\overline{g}}_{\alpha m n}\!(\tau_1,\tau_2)$ in matrix
form can be written as
\begin{eqnarray}
	\label{g02}
	{\bf g}_{\alpha} (t,\tau) = \int \frac{d\epsilon}{2\pi} ~ {\bf J}_{\alpha}(\epsilon)  e^{-i \epsilon (t-\tau)},
\end{eqnarray}
\vskip -0.5cm
\begin{eqnarray}
	\label{gtilde2}
	{\widetilde{\bf g}}_{\alpha} (\tau_1,\tau_2) = \int \frac{d\epsilon}{2\pi} ~ {\bf J}_{\alpha}(\epsilon) f_{\alpha}(\epsilon)
	e^{-i \epsilon (\tau_1-\tau_2)},
\end{eqnarray}
\vskip -0.5cm
\begin{eqnarray}
	\label{gtbar2}
	{\overline{\bf g}}_{\alpha} (\tau_1,\tau_2) = \int \frac{d\epsilon}{2\pi} ~ {\bf J}_{\alpha}(\epsilon) \left(1- f_{\alpha}(\epsilon) \right)
	e^{-i \epsilon (\tau_1-\tau_2)},
\end{eqnarray}
where $J_{\alpha i j}(\epsilon) $
is the spectral density. Here $f_{\alpha}(\epsilon)=1/[e^{\beta_\alpha (\epsilon -\mu_\alpha)} + 1]$ is the Fermi-Dirac
distribution of electrode $\alpha$ at time $t_0$ with the chemical potential $\mu_\alpha$ and initial
reservoir temperature $\beta_\alpha=1/k_B T_\alpha$. We assume the Lorentzian line shape
\cite{MatisseTu2008non,jin2010non,yang2014transient} associated to the electronic structure of the electrodes as
\begin{eqnarray}
	J_{\alpha ij}(\epsilon) = \frac{\Gamma^{\alpha}_{ij} W_\alpha^2}{(\epsilon-\mu_\alpha)^2 + W_\alpha^2},
\end{eqnarray}
where $W_\alpha$ is the bandwidth of the Lorentzian spectral distribution.
For the dots being in series (see Fig.~\ref{fig1a}), $\Gamma^{L}_{11}=\Gamma^L$, $\Gamma^{R}_{22}=\Gamma^R$,
$\Gamma^{L}_{22}=\Gamma^{R}_{11}=0$, and $\Gamma^{\alpha}_{12}=\Gamma^{\alpha}_{21}=0$.
When the dots are in parallel, $\Gamma^{\alpha}_{11}=\Gamma^{\alpha}_{22}=\Gamma^{\alpha}/2$
and $\Gamma^{\alpha}_{12}=\Gamma^{\alpha}_{21}= \sqrt{\Gamma^{\alpha}_{11} \Gamma^{\alpha}_{22} }$, as shown in Fig.~\ref{fig1b}.

\subsection{Two-time corelation functions for LGI correlators}

The two-time correlation function of the observable $Q(t)$ is given by
\begin{eqnarray}
\nonumber
&& \!\!\!\!\! \langle Q(t_2)  Q(t_1) \rangle \\
&&{} \!\!\!\!\!= 4 \langle n_2(t_2)  n_2(t_1) \rangle - 2 \langle n_2(t_2) \rangle
- 2 \langle n_2(t_1) \rangle + 1.
\label{C21t}
\end{eqnarray}
One can show that for an initial state $|0 1\rangle$, \textit{i.e} initially the first dot is unoccupied and
the second dot is occupied, the exact two-time correlation function $\langle n_2(t_2)  n_2(t_1) \rangle$ can
be expressed in terms of the nonequilibrium Green's functions $u_{ij}(t,t_0)$ and $v_{ij}(t_1,t_2 )$ and
${\overline v}_{ij}(t_1,t_2)$ as
\begin{eqnarray}
\nonumber
&& \!\!\!\!\! \langle n_2(t_2)  n_2(t_1) \rangle=\langle a_2^{\dag}(t_2)  a_2(t_2) a_2^{\dag}(t_1)  a_2(t_1) \rangle \\
\nonumber
&&{} \!\!\!\!\!=u_{22}^{\ast}(t_2) u_{22}(t_1) u_{21}(t_2) u_{21}^{\ast}(t_1) + \big|u_{22}(t_2)\big|^2 \big|u_{22}(t_1)\big|^2 \\
\nonumber
&&{} + \big|u_{22}(t_2)\big|^2 v_{22}(t_1,t_1) + \big|u_{22}(t_1)\big|^2 v_{22}(t_2,t_2) \\
\nonumber
&&{} + u_{21}(t_2) u_{21}^{\ast}(t_1) v_{22}(t_1,t_2) + u_{22}^{\ast}(t_2) u_{22}(t_1) {\overline v}_{22}^{\ast}(t_1,t_2) \\
&&{} + v_{22}(t_2,t_2) ~v_{22}(t_1,t_1) + v_{22}(t_1,t_2) ~{\overline v}_{22}^{\ast}(t_1,t_2),
\label{n2n1}
\end{eqnarray}
where the noise correlation functions $v_{ij}(t_1,t_2)$$=\langle F_{j}^\dagger(t_2) F_{i} (t_1) \rangle$ and
${\overline v}_{ij}(t_1,t_2)$$=\langle F_{i} (t_1) F_{j}^\dagger(t_2) \rangle$ are evaluated using the solution
of Eq.~(\ref{noise}). See Eqs.~(\ref{vij}) and (\ref{vijbar}) for the detailed expressions of the correlation functions.
The function $v_{ij}(t_1,t_2)$ is related to the lesser Green function in Keldysh formalism
\cite{jin2010non}. If we assume that the central dot system and the electrodes are initially decoupled at time $t_0$,
and the electrodes are initially in thermal equilibrium, then using the solution of Eq.~(\ref{noise}) with the initial
condition $F_{i}(t_0)=0$ one can evaluate the noise correlation functions.
The two-time correlation functions (\textit{i.e.} Green's functions: $u_{i,j}(t,t_0)$, $v_{ij}(t_1,t_2)$) being experimentally measurable are
central to the understanding of a wide range of non-equilibrium and statistical phenomena for studying quantum many-body systems \cite{haug1996,Imry2002}. In quantum transport, the two-time correlation functions of the
electric current through nanostructure devices are utilised to analyse the noise spectrum and current fluctuations
\cite{feng2008current,clerk2010introduction,yang2014transient,thibault2015pauli}.
For open quantum systems, the system-environment back-action processes that disclose the
non-Markovian memory effects are revealed by two-time correlation functions that link a past event with
its future, and the non-Markovianity has been measured using two-time correlation functions \cite{ali2015non}.
The single-time average values $\langle n_2(t_1) \rangle$ and $\langle n_2(t_2) \rangle$ with respect to the
double dot initial state $|01\rangle$ are respectively given by
\begin{eqnarray}
\label{n2t1}
\langle n_2(t_1) \rangle &=& \langle a_2^{\dag}(t_1)  a_2(t_1) \rangle = |u_{22}(t_1)|^2 + v_{22} (t_1,t_1), \\
\label{n2t2}
\langle n_2(t_2) \rangle &=& \langle a_2^{\dag}(t_2)  a_2(t_2) \rangle = |u_{22}(t_2)|^2 + v_{22} (t_2,t_2).
\end{eqnarray}
Finally, considering the local charge detection at four different time $t_1=0$,  $t_2= t_1 + \tau$, $t_3= t_1+2\tau$, and
$t_4= t_1+3\tau$ where $\tau$ is the time interval between the measurements, we calculate the correlation terms
$C_{21}$, $C_{32}$, $C_{31}$, $C_{43}$, and $C_{41}$ for this double quantum dot system to finally
estimate the value of $C_3$ and $C_4$. Note that the two-time operators $Q(t_j) Q(t_i)$ are not Hermitian in general,
for which the correlation function $C_{ji}=\langle Q(t_j) Q(t_i) \rangle$ can be a complex quantity. We take symmetric
combinations $\langle \{ Q(t_j) Q(t_i) \} \rangle/2$ to identify them with physical expectation values of the two-time
measurements. The symmetrised operators $(Q(t_j) Q(t_i) + Q(t_i) Q(t_j))/2$ are Hermitian, whose expectation
values provide real average values of the two-time measurements \cite{fritz2010quantum,emary2014leggett,chen2014investigating,ali2017probing}.

\begin{figure*}[htp]
\centering
\begin{subfigure}[b]{0.45\textwidth}
\centering
\includegraphics[width=\textwidth]{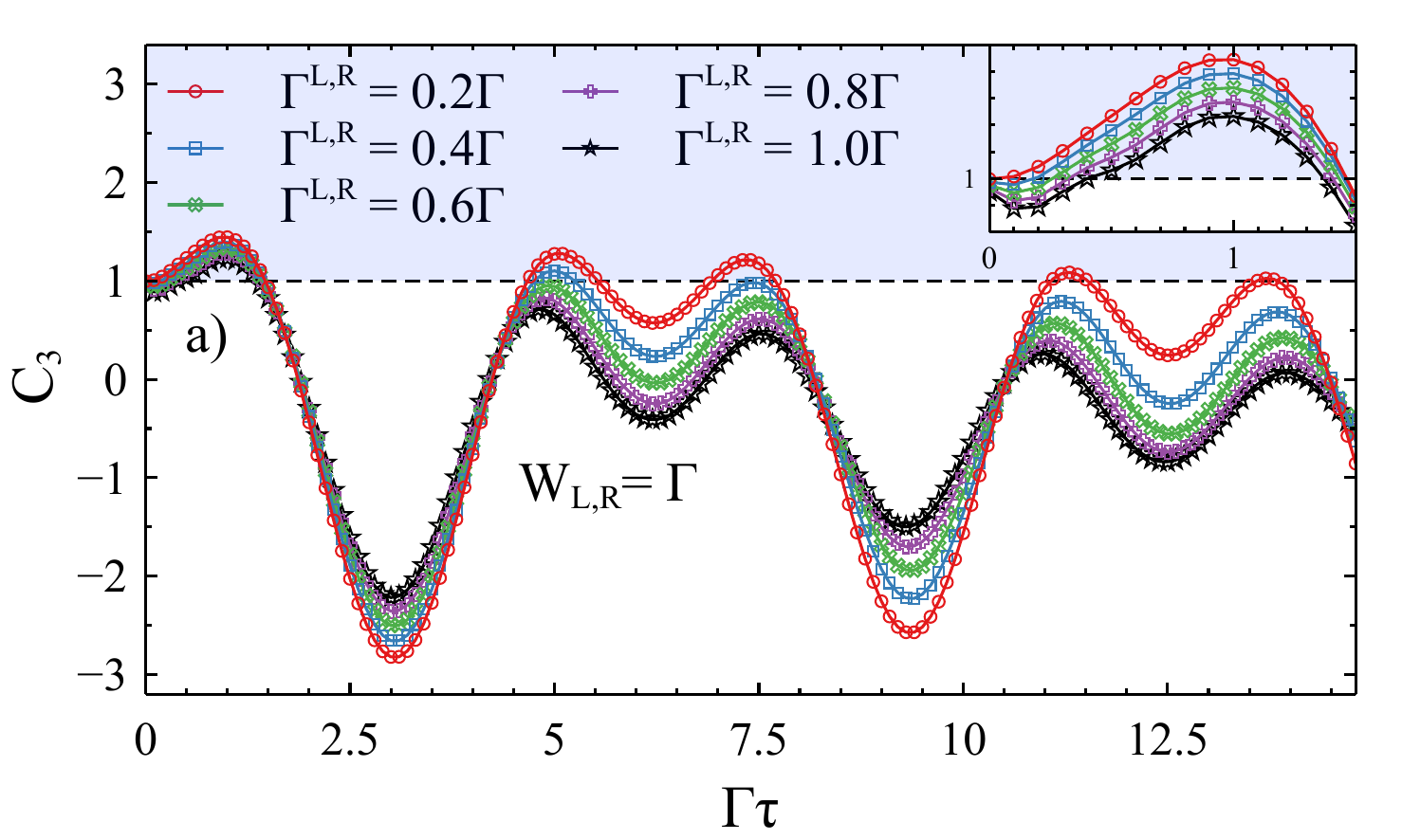}
\end{subfigure}
\hskip 0.0cm
\begin{subfigure}[b]{0.45\textwidth}
\centering
\includegraphics[width=\textwidth]{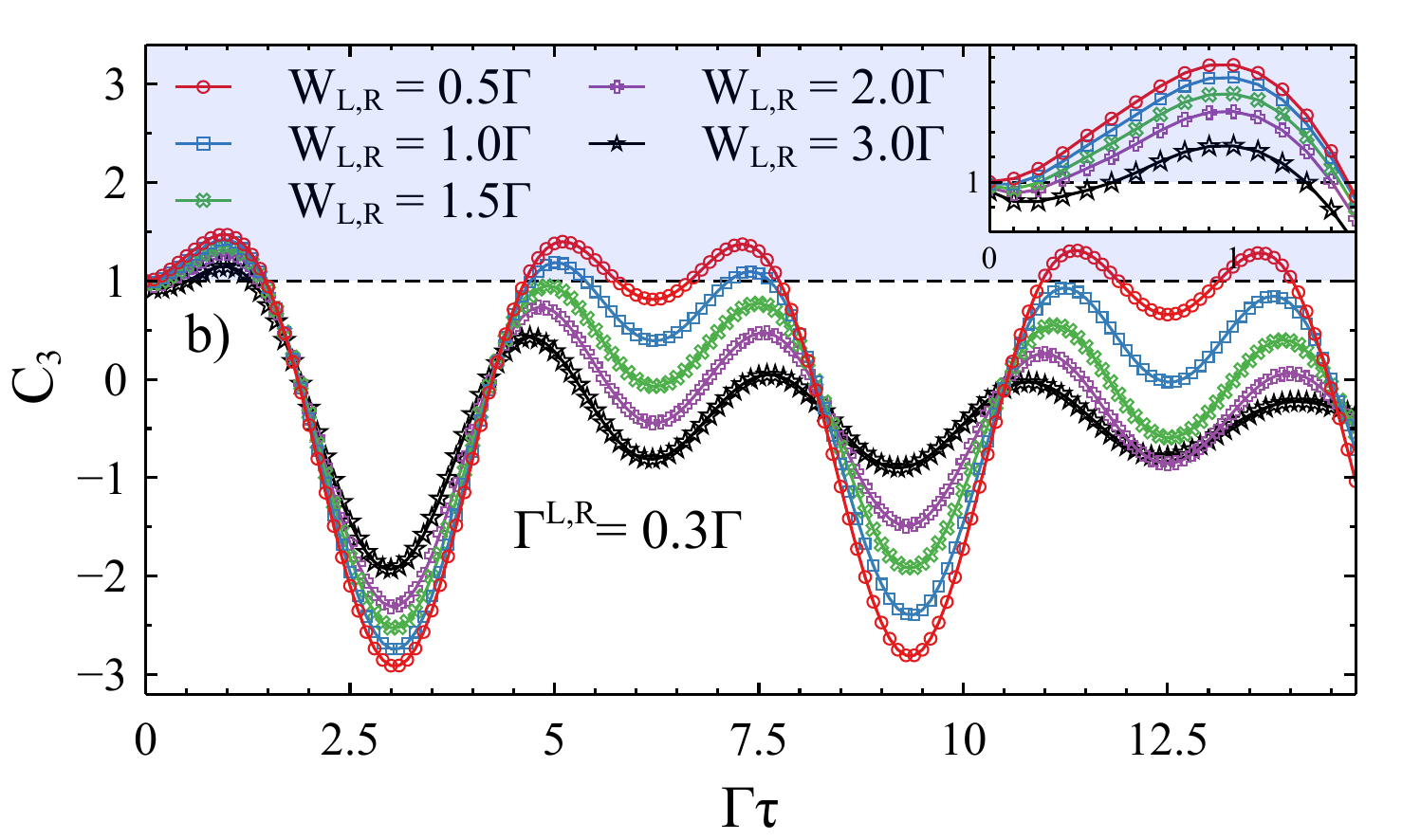}
\end{subfigure}
\caption{Exact dynamics of Leggett-Garg inequality correlator $C_3$ is shown in presence of the electronic
reservoirs by (a) varying the coupling strengths $\Gamma^L$ and $\Gamma^R$ with fixed spectral bandwidths
$W_L=W_R=\Gamma$ (b) varying the spectral bandwidth $W_L$ and $W_R$ with fixed coupling strengths
$\Gamma^L=\Gamma^R=0.3\Gamma$. The other parameter values are taken as $\mu_L=5\Gamma$,
$\mu_R=-5\Gamma$, and the temperature $k_B T_L=k_B T_R=0.1\Gamma$.}
\label{fig4}
\end{figure*}

\begin{figure*}[htp]
	\centering
	\begin{subfigure}[b]{0.45\textwidth}
		\centering
		\includegraphics[width=\textwidth]{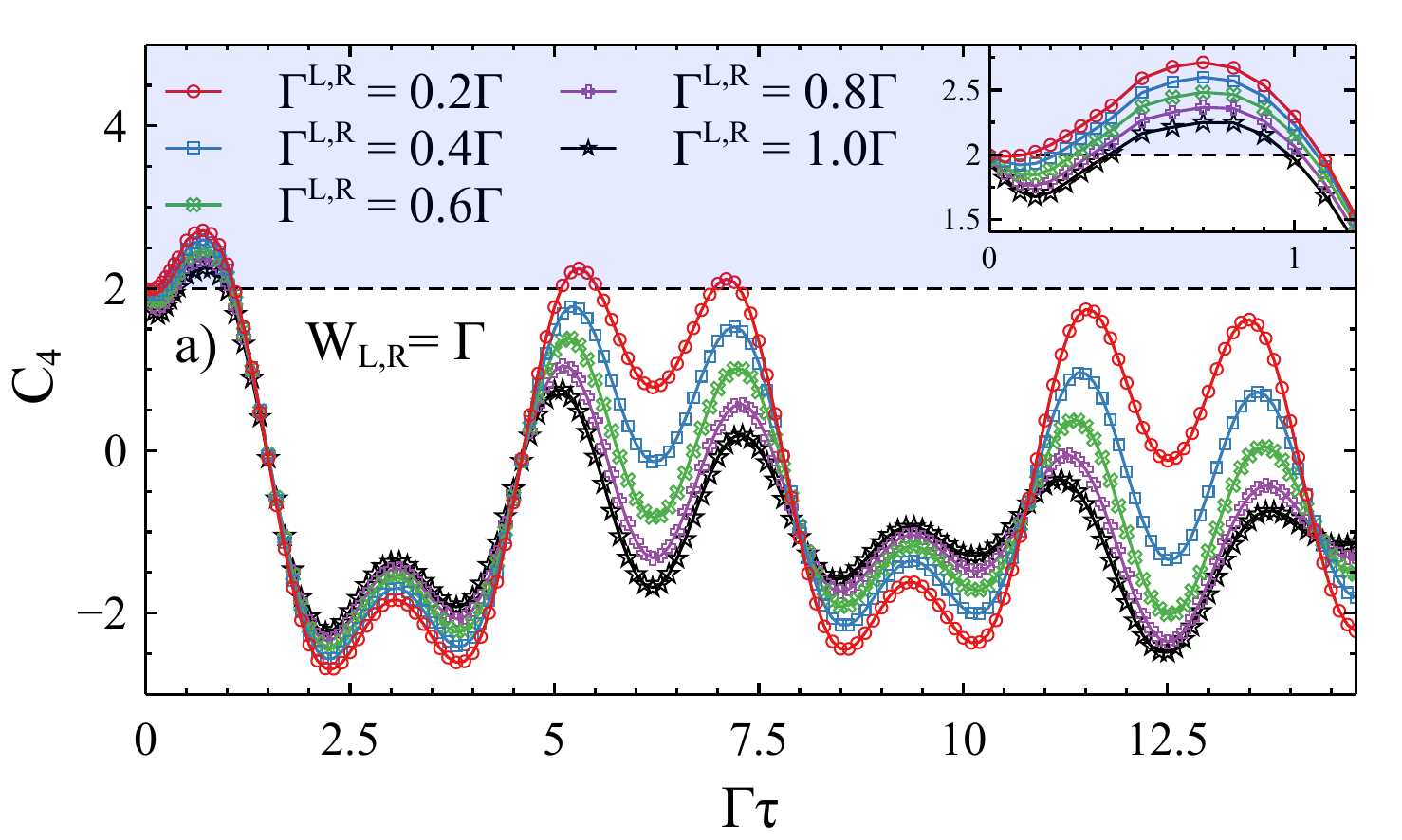}
	\end{subfigure}
	\hskip 0.0cm
	\begin{subfigure}[b]{0.45\textwidth}
		\centering
		\includegraphics[width=\textwidth]{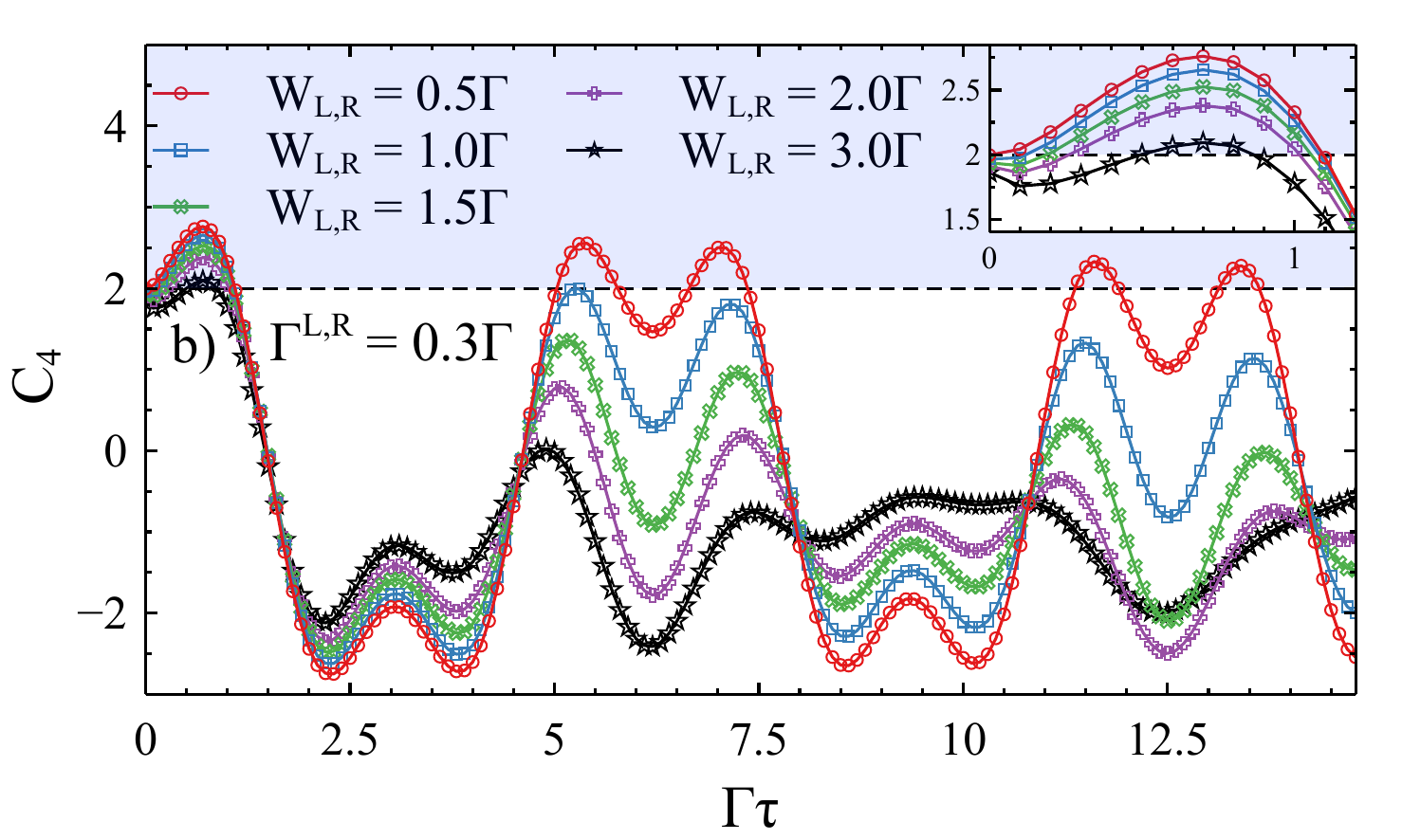}
	\end{subfigure}
	\caption{Exact dynamics of Leggett-Garg inequality correlator $C_4$ is shown in presence of the electronic
		reservoirs by (a) varying the coupling strengths $\Gamma^L$ and $\Gamma^R$ with fixed spectral bandwidths
		$W_L=W_R=\Gamma$ (b) varying the spectral bandwidth $W_L$ and $W_R$ with fixed coupling strengths
		$\Gamma^L=\Gamma^R=0.3\Gamma$. The other parameter values taken as $\mu_L=5\Gamma$,
		$\mu_R=-5\Gamma$, and the temperature $k_B T_L=k_B T_R=0.1\Gamma$.}
	\label{fig5}
\end{figure*}

\subsection{Violation of LGI for a serially coupled quantum dot device}

We consider the situation when the dots are coupled to the electrodes. Since our system is coupled to fermionic reserviors, it can be understood as a open quantum system. The dynamics of the system is now governed
by the quantum Langevin equation (\ref{Langevin}). In the case of open quantum devices, the interaction
with the fermionic reservoirs introduces environmental noise, leading to decoherence and dissipation, thereby
losing the system's quantumness or non-classical behavior. First we consider that the dots are in a series configuration
as shown in Figure (\ref{fig1a}). The initial state of the double quantum dot system is taken as $ \vert 01 \rangle$. The
double dot energy levels are taken as $\epsilon_{11}=\epsilon_{22}=\Gamma$. The off-diagonal
tunneling term in the dot Hamiltonian $\epsilon_{12}=\epsilon_{21}=0.5\Gamma$.

In Fig.~(\ref{fig4}a), we show the exact dynamics of Leggett-Garg inequality correlator $C_3$ as a function
of $\tau$ for different system-reservoir coupling strengths $\Gamma^L$ and $\Gamma^R$. The spectral
bandwidths of the left and right electrodes are fixed at $W_L=W_R=\Gamma$. The chemical potentials and
temperatures of the reservoirs are taken as $\mu_L=5\Gamma$, $\mu_R=-5\Gamma$ and
$k_B T_L=k_B T_R=0.1\Gamma$. Different curves represent different
coupling strengths, namely $\Gamma^L=\Gamma^R=0.2\Gamma$ (red), $\Gamma^L=\Gamma^R=0.4\Gamma$ (blue),
$\Gamma^L=\Gamma^R=0.6\Gamma$ (green), $\Gamma^L=\Gamma^R=0.8\Gamma$ (violet),
$\Gamma^L=\Gamma^R=\Gamma$ (black). The double quantum dot system shows quantum behavior
(violation of LGI with $C_3 > 1$) in the short $\tau$ regime for all these coupling strengths.
We see quantum violation of LGI (\ref{LGI1}) in long $\tau$ only when the system-reservoir coupling
strength is relatively weak ($\Gamma^L=\Gamma^R=0.2\Gamma$). The long-time violation of LGI vanishes
for higher values of coupling strengths. In case of weak coupling ($\Gamma^L=\Gamma^R=0.2\Gamma$), the central system behaves more like a isolated double dot system and thus it retains more quantumness in longer time interval and the violation persists for higher value of $\tau$. In case of strong coupling, the influence of the reservoir on the central system is more prominent and the central system is driven towards the non equlibrium steady state faster. Thus decoherence beomces more prominent as time goes on making the cenral system lose its quantum properites, there by giving no violation for larger values of $\tau$.

Next in Fig.~(\ref{fig4}b), we show the exact dynamics of $C_3$ by varying the spectral bandwidths
$W_L$ and $W_R$ of the left and right reservoirs. Fixed values of coupling strengths are symmetrically
taken as $\Gamma^L=\Gamma^R=0.3\Gamma$. We vary the Lorentzian level broadening parameters as
$W_L=W_R=0.5\Gamma$ (red), $W_L=W_R=\Gamma$ (blue), $W_L=W_R=1.5\Gamma$ (green),
$W_L=W_R=2\Gamma$ (violet), $W_L=W_R=3\Gamma$ (black).
The system dynamics goes beyond classical description (violation of LGI) for short measurement
intervals $\tau$, which we see for all the spectral widths considered here. Quantum violation of LGI
(\ref{LGI1}) is obtained in long $\tau$ only when the spectral width of the reservoirs take small values
($W_L=W_R< \Gamma$). Long-time violation of LGI vanishes as one increases the spectral widths of
the reservoirs. This indicates that the non-Markovian memory effects of the electronic reservoirs
play an important role in obtaining the LGI violation, and we do not see any violation of LGI in the
broadband limit. We also confirm these observations through the dynamics of LGI correlator $C_4$ in
Figs.~(\ref{fig5}a) and (\ref{fig5}b).

\subsection{Violation of LGI for a parallel-coupled quantum dot device}

\begin{figure*}[h]
	\centering
	\begin{subfigure}[h]{0.45\textwidth}
		\centering
		\includegraphics[width=\textwidth]{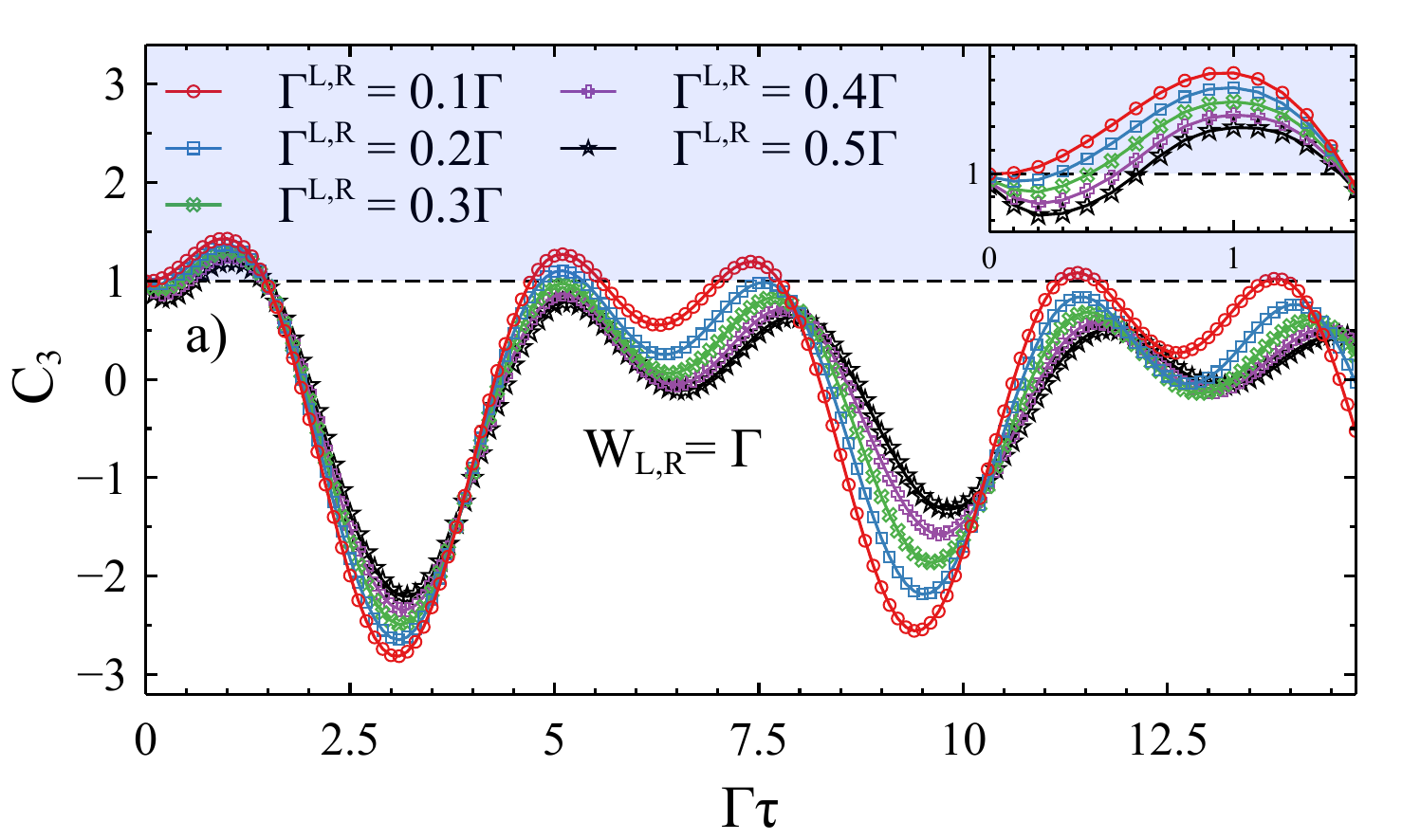}
	\end{subfigure}
	\hskip 0.0cm
	\begin{subfigure}[h]{0.45\textwidth}
		\centering
		\includegraphics[width=\textwidth]{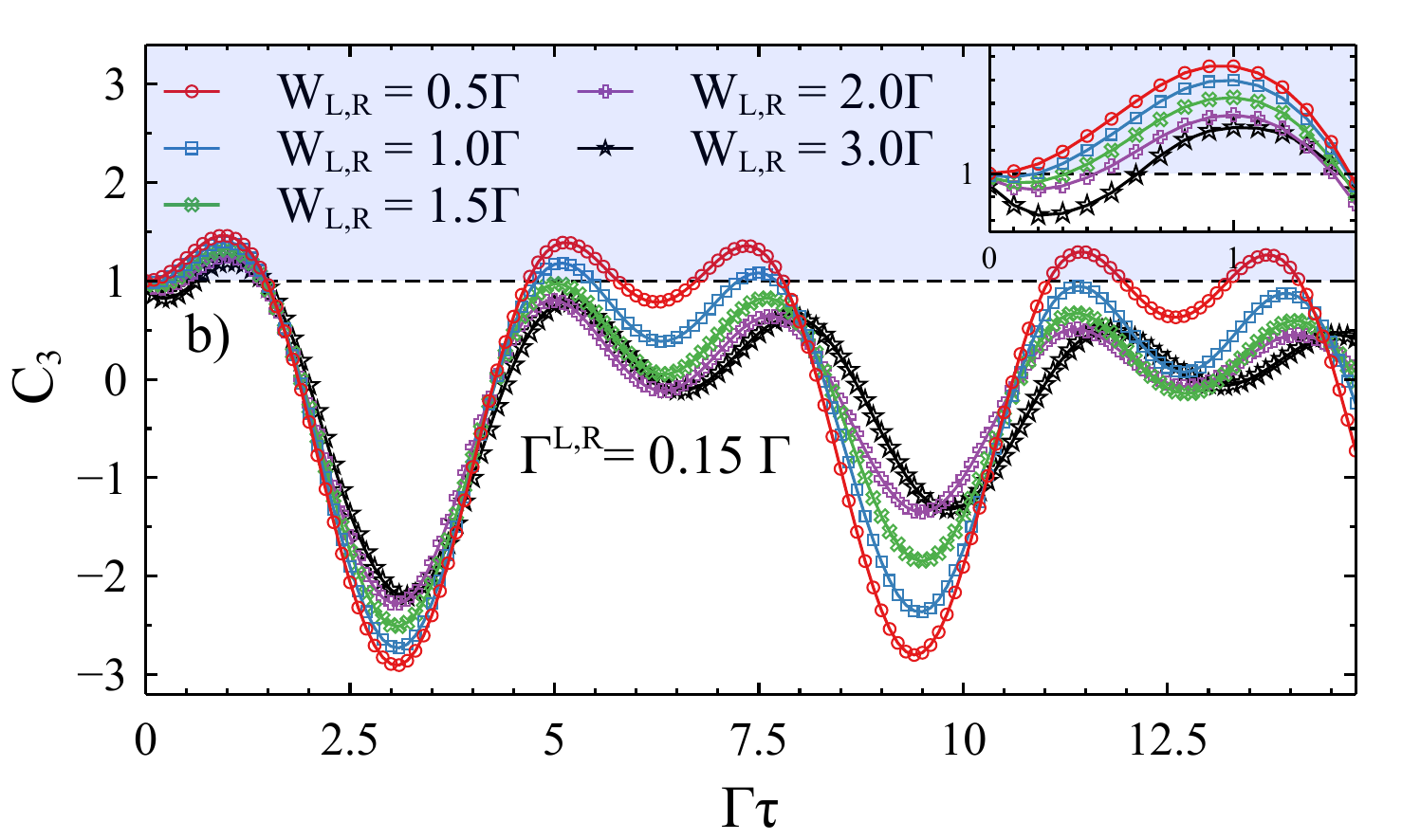}
	\end{subfigure}
	\caption{Exact dynamics of Leggett-Garg inequality correlator $C_3$ is shown in presence of the electronic
		reservoirs by (a) varying the coupling strengths $\Gamma^{L}_{11}$, $\Gamma^{L}_{22}$, $\Gamma^{R}_{11}$   and $\Gamma^{R}_{22}$ with fixed spectral bandwidths $W_L=W_R=\Gamma$ (b) varying the spectral bandwidth
		$W_L$ and $W_R$ with fixed coupling strengths $\Gamma^{L}_{11}=\Gamma^{L}_{22}=\Gamma^{R}_{11}=\Gamma^{R}_{22}=0.15\Gamma$.
		The other parameter values are taken as $\mu_L=5\Gamma$,
		$\mu_R=-5\Gamma$, $k_B T_L=0.1\Gamma$, $k_B T_R=0.1\Gamma$.}
	\label{fig6}
\end{figure*}

We next consider the situation when two dots are parallelly coupled with the left and right electronic reservoirs as
shown in Fig.~1b. In Fig.~(\ref{fig6}a), we show the exact dynamics of $C_3$ as a function of
$\tau$ for different system-reservoir coupling strengths $\Gamma^{L}_{11}$, $\Gamma^{L}_{22}$, $\Gamma^{R}_{11}$   and $\Gamma^{R}_{22}$.
We have fixed the spectral bandwidths of the left and right electrodes at $W_L=W_R=\Gamma$.
The chemical potentials and temperatures of the reservoirs are taken as $\mu_L=5\Gamma$, $\mu_R=-5\Gamma$ and
$k_B T_L=k_B T_R=0.1\Gamma$. Different curves represent different
coupling strengths, namely $\Gamma^{L}_{11}=\Gamma^{L}_{22}=\Gamma^{R}_{11}=\Gamma^{R}_{22}=0.1\Gamma$ (red),
$0.2\Gamma$ (blue), $0.3\Gamma$ (green), $0.4\Gamma$ (violet), and $0.5\Gamma$ (black).
For this parallel configuration of the double quantum dot system we also see quantum behavior
(violation of LGI with $C_3 > 1$) in the short $\tau$ regime for all these coupling strengths.
We see quantum violation of LGI in long $\tau$ only when the system-reservoir coupling
strength is relatively weak ($\Gamma^{L}_{11}=\Gamma^{L}_{22}=\Gamma^{R}_{11}=\Gamma^{R}_{22}=0.1\Gamma$).
The long-time violation of LGI vanishes for higher values of coupling strengths.

\begin{figure*}[h]
	\centering
	\begin{subfigure}[h]{0.45\textwidth}
		\centering
		\includegraphics[width=\textwidth]{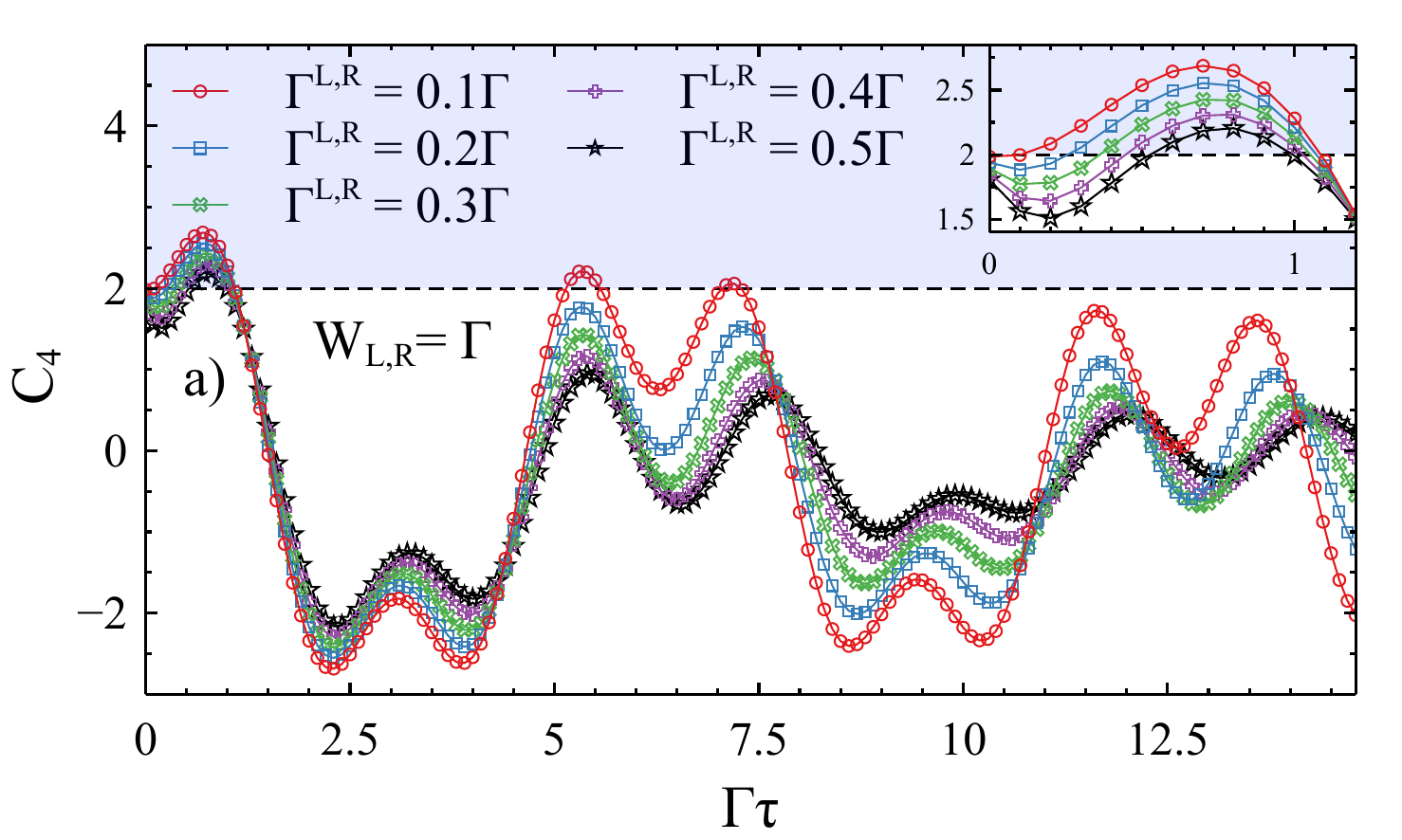}
	\end{subfigure}
	\hskip 0.0cm
	\begin{subfigure}[h]{0.45\textwidth}
		\centering
		\includegraphics[width=\textwidth]{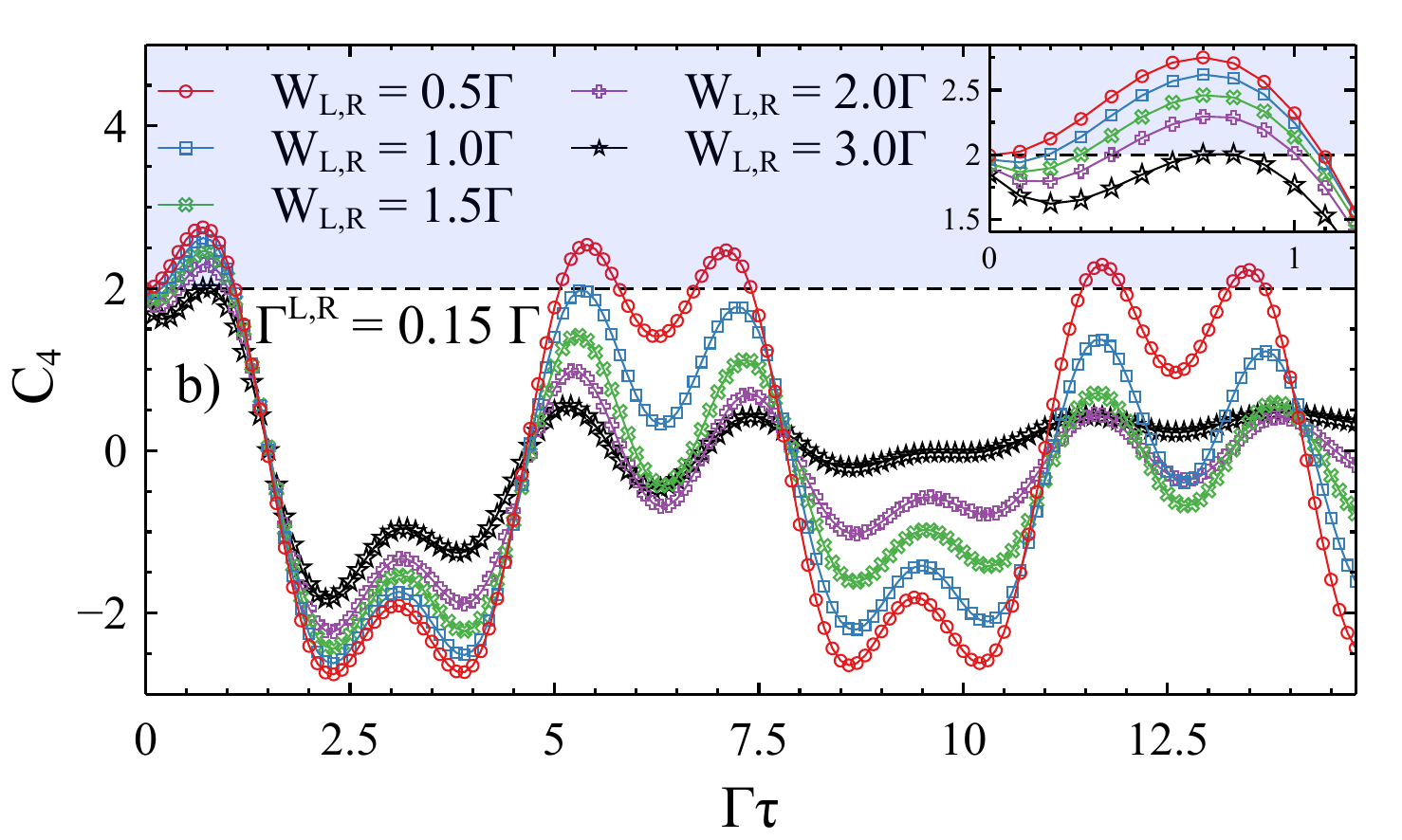}
	\end{subfigure}
	\caption{Exact dynamics of Leggett-Garg inequality correlator $C_4$ is shown in presence of the electronic
		reservoirs by (a) varying the coupling strengths $\Gamma^{L}_{11}$, $\Gamma^{L}_{22}$, $\Gamma^{R}_{11}$   and $\Gamma^{R}_{22}$ with fixed spectral bandwidths
		$W_L=W_R=\Gamma$ (b) varying the spectral bandwidth $W_L$ and $W_R$ with fixed coupling strengths
		$\Gamma^{L}_{11}=\Gamma^{L}_{22}=\Gamma^{R}_{11}=\Gamma^{R}_{22}=0.15\Gamma$.
		The other parameter values taken as $\mu_L=5\Gamma$,
		$\mu_R=-5\Gamma$, and the temperature $k_B T_L=k_B T_R=0.1\Gamma$.}
	\label{fig7}
\end{figure*}

Next in Fig.~(\ref{fig6}b), we show the exact dynamics of $C_3$ by varying the spectral bandwidths
$W_L$ and $W_R$ of the left and right reservoirs. Fixed values of coupling strengths are symmetrically
taken as $\Gamma^{L}_{11}=\Gamma^{L}_{22}=\Gamma^{R}_{11}=\Gamma^{R}_{22}=0.15\Gamma$.
We vary the Lorentzian level
broadening parameters as $W_L=W_R=0.5\Gamma$ (red), $\Gamma$ (blue), $1.5\Gamma$ (green),
$2.0\Gamma$ (violet), and $3.0\Gamma$ (black). For the above values of spectral widths, the
system dynamics goes beyond classical regime (violation of LGI) for short measurement
intervals $\tau$. Quantum violation of LGI is obtained in long $\tau$ only when the spectral width
of the reservoirs take small values ($W_L=W_R< \Gamma$). Long-time violation of LGI vanishes
as one increases the spectral widths of the reservoirs. This indicates that the non-Markovian memory
effects of the electronic reservoirs play an important role in obtaining the LGI violation, and we do
not see any violation of LGI in the broadband limit. For this parallel configuration of the dots
we also confirm similar observations through the dynamics of LGI correlator $C_4$ in
Figs.~(\ref{fig7}a) and (\ref{fig7}b).
\begin{figure}[h]
\includegraphics[width=0.5\columnwidth]{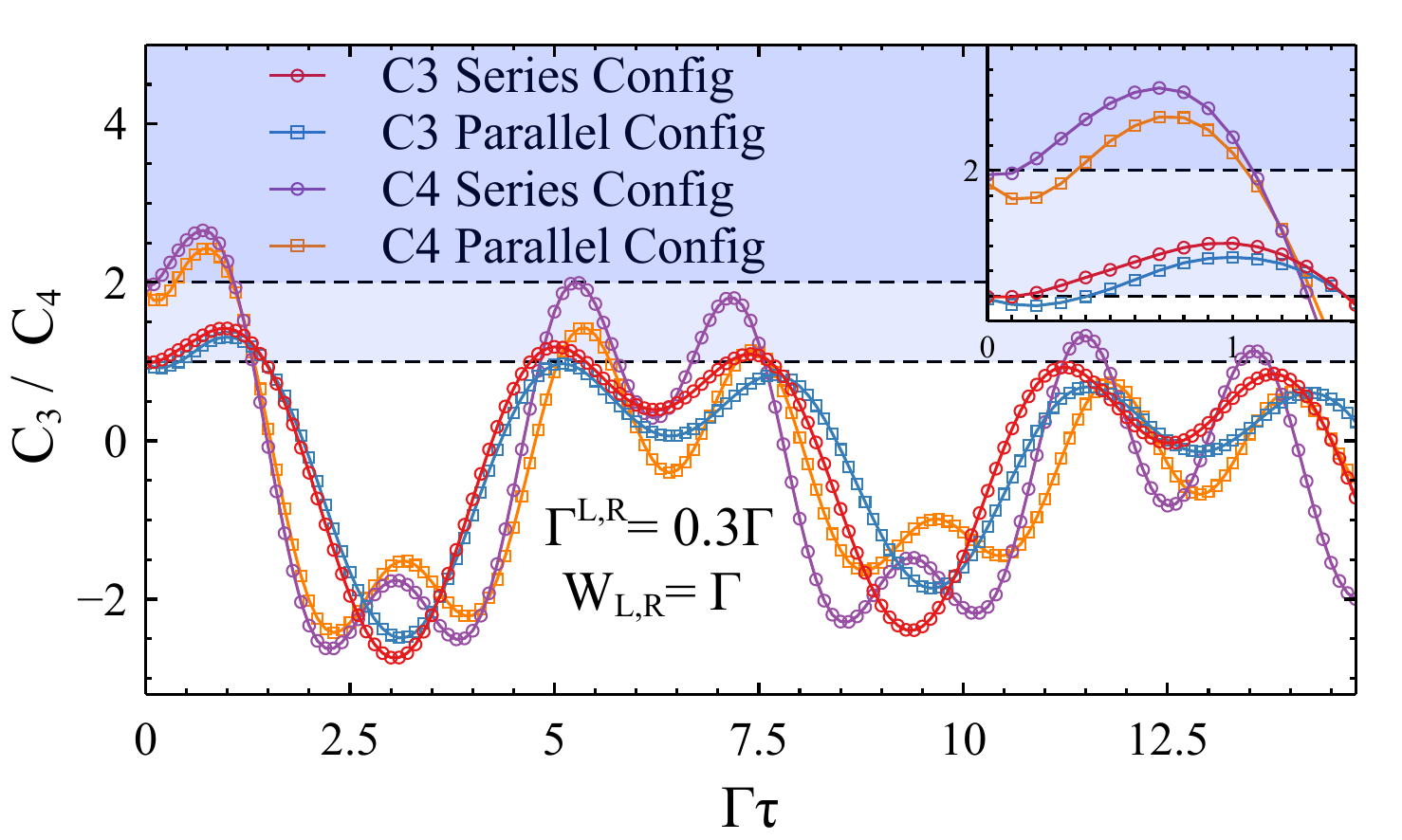}
\caption{Comparative plot of LGI for series and parallel configurations. Exact dynamics of $C_3$ and $C_4$ are
shown with coupling strengths $\Gamma^{L,R}=0.3\Gamma$. The spectral bandwidths are taken as
$W_L=W_R=\Gamma$.}
\label{fig8}
\end{figure}

In the parallel configuration, both dots connect to both reservoirs, while in the series configuration, each dot connects to only one reservoir.
Therefore, we set $\Gamma^{L,R}=0.15\Gamma$ for the parallel and $\Gamma^{L,R}=0.3\Gamma$ for the series
configuration. For a strict comparison, we now use $\Gamma^{L,R}=0.3\Gamma$ for both configurations and present the
dynamics of $C_3$ and $C_4$ in Fig.~\ref{fig8}. The parallel configuration shows less violation of LGI compared to the
series configuration, as the detrimental effect of the reservoirs is more pronounced in the parallel setup.

\section{Conclusions}\label{sec:results}

Quantum nanostructure devices have generated much interest due to their potential applications in developing quantum technologies, where the quantum coherence of electrons is the prime ingredient. Artificial atoms and molecules can now be designed with semiconductor quantum dots thanks to advances in quantum nanofabrication technology. Single or double quantum dot devices are used to manipulate electronic states coherently \cite{heinrich2021quantum,tarucha1996shell,oosterkamp1998microwave,fujisawa1998spontaneous,
van2002electron,hayashi2003coherent,petta2005coherent,brandes2005coherent}. In the presence of electronic reservoirs,
quantum dot devices may well constitute what is known as open quantum systems \cite{carmichael1999statistical,breuer2002theory,gardiner2004quantum}. The quantum master equation method is
used to study non-equilibrium transport across these nanostructures \cite{MatisseTu2008non,jin2010non,tu2012transient},
where the system loses its quantumness due to dissipation and decoherence brought by environmental noise. However, it is often
challenging to know the extent to which these nanoelectronic devices can behave quantum mechanically, and how to tune the system-reservoir parameters to make them work in the quantum regime. Recent experimental investigations on double quantum dot systems
\cite{kiesslich2007noise,shinkai2009correlated,kagan2015charge,dorsch2021heat} provide added motivation to study
the LGI violation in such systems. Probing the nonclassical or quantum nature of these nanoelectronic devices remains therefore of great significance in the emerging world of quantum technologies. In this work, we employ Leggett Garg inequalities (LGI) to test the quantumness
of electron transport in nanoelectronic devices, where the nonequilibrium Green's functions determine the temporal LGI correlators.
Many of the previous studies on the LGI violation of the open systems relied on the Born-Markov approximation, where the memory effect was completely ignored. However, the study of non-Markovian dynamics of quantum systems interacting with reservoirs becomes significant for understanding the relaxation phenomenon in the ultrafast transient regime to mimic what happens to high-speed quantum devices. In light of this, our results apply to both Markovian and non-Markovian regimes to encompass an experimental situation where the devices can have strong coupling with the electronic reservoirs. Our detailed analysis is also able to capture the effect of finite reservoir correlation time by accounting for level-broadening at the electrodes and non-Markovian memory effects. Further, the large bias restriction is no longer imposed in our calculations so that we can comfortably consider a finite bias between the electronic reservoirs. Here we make use of the Heisenberg equation of motion approach to obtain the exact dynamics of the two-time correlation functions in terms of non-equilibrium Green’s functions. The two-time correlation functions, which are experimentally measurable, can often play a crucial role in the complex dynamics of dissipative many-body quantum systems and also in the transient quantum transport to study the current fluctuations and noise spectrum
\cite{feng2008current,clerk2010introduction,yang2014transient} as the system goes out of equilibrium \cite{knezevic2003memory,stefanucci2007bound,sciolla2015two,thibault2015pauli}. Our approach may open up some possibilities of witnessing the quantumness for other quantum many-body systems as well that are somehow driven out of equilibrium.

\section*{Acknowledgments}
TYM, SK, and AS would like to thank the Science and Engineering Research Board (SERB), Government of India \textit{via} Project No. CRG/2022/007836 for the necessary financial support, and also, for providing  computational resources  through SERB-funded GPGPU system at SRMIST, India.
\appendix

\section{Derivation of LGI}\label{sec:appA}

Leggett-Garg inequality (\ref{LGI1}) can be constructed by adopting two classical assumptions: (a) measurement
on a classical system reveals a well-defined pre-existing value and (b) that value can be measured without
disturbing the system. We consider a classical dichotomic variable $Q(t)$ which can take values $+1$ or $-1$
whenever measured. The measurement value of the observable at time $t_i$ is denoted as $Q(t_i)=Q_i$. One
can perform three set of experimental runs where in the first set of runs the observable $Q(t)$ is measured at times
$t_1$ and $t_2$; in the second run, $Q(t)$ is measured at $t_1$ and $t_3$; and in the third run $Q(t)$ is measured
at $t_2$ and $t_3$ (where $t_3 > t_2 > t_1 $). The classical assumptions mentioned above imply the existence of a
joint probability distribution \cite{leggett1985quantum,emary2014leggett,das2014unification,markiewicz2014unified}
to describe the time-separated measurement statistics of all three experimental runs. The correlation
function $C_{ji}$ is obtained from the joint probability $P_{ji}(Q_j,Q_i)$ of obtaining the results $Q_i=Q(t_i)$
and $Q_j=Q(t_j)$ from measurements at times $t_i$, $t_j$ as
\begin{eqnarray}
C_{ji} = \sum_{Q_j,Q_i=\pm 1} Q_j Q_i P_{ji}(Q_j,Q_i).
\label{Cij}
\end{eqnarray}
The subscripts on P indicate the times at which the measurements were made. According to the
assumption (a), the two-time probability
can be obtained as the marginal of a three-time probability distribution (since
observable $Q$ has always a well-defined value, even in absence of measurement):
\begin{eqnarray}
P_{ji}(Q_j,Q_i) = \sum_{Q_k=\pm 1} P_{kji}(Q_k,Q_j,Q_i),
\label{Pij}
\end{eqnarray}
with $k \ne i,j$. Assumption (b) implies that measurements do not affect the state of the system or the subsequent dynamics.
Under this assumption, one can drop the subscripts of $P_{ji}$ and $P_{kji}$ as the time indices in $P$ are not important.
One can then use a single joint probability distribution $P(Q_k,Q_j,Q_i)$ to calculate all the correlation functions, namely
$C_{21}$, $C_{32}$, and $C_{31}$ as
\begin{eqnarray}
\nonumber
&& \!\!\!\!\!\!\!\!C_{21}\!=\!P(+,+,+)\!+\!P(-,+,+)\!+\!P(+,-,-)\!+\!P(-,-,-) \\
&&{} -P(+,+,-)\!-\!P(-,+,-)\!-\! P(+,-,+)\!-\!P(-,-,+),
\label{C21a}
\end{eqnarray}
\vskip -0.9cm
\begin{eqnarray}
\nonumber
&& \!\!\!\!\!\!\!\!C_{32}\!=\!P(+,+,+)\!+\!P(+,+,-)\!+\!P(-,-,+)\!+\!P(-,-,-) \\
&&{} -P(+,-,+)\!-\!P(+,-,-)\!-\! P(-,+,+)\!-\!P(-,+,-),
\label{C32a}
\end{eqnarray}
\vskip -0.9cm
\begin{eqnarray}
\nonumber
&& \!\!\!\!\!\!\!\!C_{31}\!=\!P(+,+,+)\!+\!P(+,-,+)\!+\!P(-,+,-)\!+\!P(-,-,-) \\
&&{} -P(+,+,-)\!-\!P(+,-,-)\!-\! P(-,+,+)\!-\!P(-,-,+),
\label{C32a}
\end{eqnarray}
\vskip -0.2cm
\noindent
where we have used the shorthand $P(\pm,\pm,\pm)=P(\pm 1,\pm 1,\pm 1)$, etc. Now using the
completeness relation
\begin{eqnarray}
\label{totP}
\sum_{Q_3,Q_2,Q_1} P(Q_3,Q_2,Q_1) = 1,
\end{eqnarray}
\vskip -0.2cm
we obtain
\vskip -0.5cm
\begin{eqnarray}
\nonumber
C_3 &=& C_{21} + C_{32} - C_{31} \\
&=& 1 - 4 \Big[ P(+,-,+) + P(-,+,-)  \Big].
\label{C3a}
\end{eqnarray}
The upper bound of $C_3$ is attained under the choice of $P(+,-,+)$ $=P(-,+,-)$ $=0$, for which
$C_3=1$. The lower bound of $C_3=-3$ can be obtained for $P(+,-,+)+P(-,+,-)=1$.

\section{Analytical solution of $w_{ij}(t,t_0)$ for the closed DQD system}\label{sec:appB}

In absence of the electronic reservoirs, the time evolution of the dot operators is given by
\begin{eqnarray}
\label{aisoln2}
a_i (t) = \sum_{j} w_{ij}(t,t_0) a_j (t_0),
\end{eqnarray}
where $w_{ij}(t,t_0)$ satisfy the following coupled differential equations
\begin{eqnarray}
\label{wij2}
\frac{d}{dt}  w_{ij}(t,t_0) =  \sum_{m} \epsilon_{im}  w_{mj}(t,t_0).
\end{eqnarray}
The analytical solutions to the above equations are as follows
\begin{eqnarray}
\nonumber
w_{11}(t,t_0) &=& A_1 \exp \Big(- \frac{i}{2} (\beta-\alpha) t \Big) \\
&& {} + A_2 \exp \Big( - \frac{i}{2} ( \beta + \alpha) t \Big), \\
\nonumber
w_{12}(t,t_0) &=& A_3 \exp \Big(- \frac{i}{2} (\beta-\alpha) t \Big) \\
&& {} + A_4 \exp \Big( - \frac{i}{2} ( \beta + \alpha) t \Big), \\
\nonumber
w_{21}(t,t_0) &=&- \frac{1}{2 \epsilon_{21}} \Big[ A_1 (\alpha - \gamma) \exp \Big(- \frac{i}{2} (\beta-\alpha) t \Big) \\
&& {} - A_2 (\alpha +\gamma)\exp \Big( - \frac{i}{2} ( \beta + \alpha) t \Big) \Big], \\
\nonumber
w_{22}(t,t_0) &=& - \frac{1}{2 \epsilon_{21}} \Big[ A_3 (\alpha - \gamma) \exp \Big(- \frac{i}{2} (\beta-\alpha) t \Big) \\
&& {} - A_4 (\alpha +\gamma)\exp \Big( - \frac{i}{2} ( \beta + \alpha) t \Big) \Big],
\end{eqnarray}
where
\begin{eqnarray}
& \alpha  = \sqrt{(\epsilon_{22}-\epsilon_{11})^2 + 4~\epsilon_{21}^2}, \nonumber \\
& \beta = \epsilon_{11} + \epsilon_{22},~~\gamma = \epsilon_{22} - \epsilon_{11}, \nonumber\\
& A_1 = \frac{\alpha + \gamma}{2 \alpha},~~A_2 = \frac{\alpha - \gamma}{2 \alpha}, \nonumber\\
& A_3 = - \frac{\epsilon_{21}}{\alpha},~~A_4 = \frac{\epsilon_{21}}{\alpha} \nonumber.
\end{eqnarray}

\bibliography{ReferenceLeggett3.bib}

\end{document}